\documentclass[10pt,twoside]{classe-cmb}
\usepackage{amssymb,amsbsy,amsmath,amsfonts,amssymb,amscd}
\usepackage{latexsym,euscript,exscale,epic,eepic,epsfig}
\usepackage[english,francais]{babel}
\usepackage{times}

\def\beq{\begin{equation}}
\def\eeq{\end{equation}}
\def\bei{\begin{itemize}}
\def\eei{\end{itemize}}
\def\bev{\begin{verbatim}}
\def\eev{\end{verbatim}}

\def\apjlett{Ap.\ J.\ Lett.}
\def\apj{Ap.\ J.}
\def\apjs{Ap.\ J.\ Supp.}
\def\prd{Phys.\ Rev.\ D}

\def\etal{et al.\ }

\def\o2{0$_{2}$}

\def\maxima{MAXIMA}

\def\boom{BOOMERanG}

\def\ruo2{RuO$_{2}$}

\def\apj{Ap.\ J.}

\def\mathrelfun#1#2{\lower3.6pt\vbox{\baselineskip0pt\lineskip.9pt
  \ialign{$\mathsurround=0pt#1\hfil##\hfil$\crcr#2\crcr\sim\crcr}}}
\def\simlt{\mathrel{\mathpalette\mathrelfun <}} 
\def\simgt{\mathrel{\mathpalette\mathrelfun >}}

\def\fun#1#2{\lower3.6pt\vbox{\baselineskip0pt\lineskip.9pt
  \ialign{$\mathsurround=0pt#1\hfil##\hfil$\crcr#2\crcr\sim\crcr}}}
\newcommand{\wisk}[1]{{\ifmmode{#1}\else{$#1$}\fi}}

\newtheorem{e-proposition}[theorem]{Proposition}

\newtheorem{e-definition}[theorem]{Definition\rm}

\ComParit{S1296-2147}
\PIT{FLA}
\PXHY{????}
\Add{?}
\Volume{0}
\Year{2003}
\FirstPage{1}
\LastPage{??}
\AuteurCourant{Stompor, Hanany, et al. }
\TitreCourant{The MAXIMA Experiment} 
\Journal
\Rubrique{Rubrique}{MAXIMA}
\SousRubrique{Sous-rubrique}{Sub-Heading}
\PresentePar{Shaul}{Hanany}
\Recu{blank}{}{}
\keywords{keyword~1~/ keyword~2~/ etc.}
\begin{document}
\selectlanguage{english}
\TitleOfDossier{The Cosmic Microwave Background}
\title{%
The MAXIMA Experiment: Latest Results and Consistency Tests
}
\author{%
R. Stompor~$^{\text{a,b}}$,
S. Hanany~$^{\text{c}}$,
M. E. Abroe~$^{\text{c}}$,
J. Borrill~$^{\text{a,b}}$,
P. G. Ferreira~$^{\text{d}}$,
A. H. Jaffe$^{\text{e}}$,
B.~Johnson$^{\text{b}}$,
A. T. Lee$^{\text{f,g}}$,
B. Rabii$^{\text{b,f}}$,
P. L. Richards$^{\text{f,g}}$,
G. Smoot$^{\text{b,f,g}}$,
C. Winant$^{\text{f}}$,
J. H. P. Wu$^{\text{h}}$
}
\address{%
\begin{itemize}\labelsep=2mm\leftskip=-5mm
\item[$^{\text{a}}$] Computational Research Division,
  Lawrence Berkeley National Laboratory, Berkeley, CA 94720, USA \\
\item[$^{\text{b}}$] Space Sciences Laboratory,
  University of California, Berkeley, CA 94720, USA \\
\item[$^{\text{c}}$] School of Physics and Astronomy, University of
  Minnesota, Minneapolis, MN 55455, USA \\
\item[$^{\text{d}}$] Astrophysics and Theoretical Physics, University 
of Oxford, UK \\
\item[$^{\text{e}}$] Astrophysics Group, Blackett Laboratory, 
Imperial College, London SW7 2BW, UK \\
\item[$^{\text{f}}$] Department 
  of Physics, University of California, Berkeley, CA 94720, USA \\
\item[$^{\text{g}}$] Physics Division,
  Lawrence Berkeley National Laboratory, Berkeley, CA 94720, USA \\
\item[$^{\text{h}}$] Department of Physics,
  National Taiwan University, Taipei, Taiwan  
\end{itemize}
}
\maketitle
\thispagestyle{empty}


\begin{Abstract}{%
The MAXIMA cosmic microwave background anisotropy experiment had a
significant impact on cosmology. Results from the program have played
a significant role in determining the geometry of the universe, given
strong supporting evidence to inflation, and, in combination with
other astrophysical data, showed that the universe is filled with dark
matter and energy.  We present a subset of the internal consistency
checks that were carried out on the \maxima-1 data prior to their
release, which demonstrate that systematics errors were much smaller
than statistical errors.  We also discuss the \maxima-2 flight and
data, compare the maps of \maxima-1 and -2 in areas where they overlap
and show that the two independent experiments confirm each other. All
of these results demonstrate that MAXIMA mapped the cosmic microwave
background anisotropy with high accuracy.  }
\end{Abstract}

\par\medskip\centerline{\rule{2cm}{0.2mm}}\medskip
\setcounter{section}{0}
\setcounter{figure}{0}
\selectlanguage{english}

\section{Introduction}

\maxima\ was a balloon-borne experiment that measured temperature
fluctuations in the cosmic microwave background (CMB) radiation.  The
instrument consisted of a 16 element array of bolometric photometers
operating between frequencies of 150 and 410 GHz. It flew twice in
1998 and 1999 from NASA's National Scientific Balloon Facility in
Palestine, Texas and the 
two flights and their corresponding data sets have become known as
\maxima-1 and \maxima-2, respectively.
Direct results such as maps and power spectra, as well as derivative
results, such as analysis techniques, cosmological implications, and
assessment of foregrounds have been published in a number of
papers. Jaffe et al.
\cite{jaffe_cmbnet} have given a compilation of \maxima\ papers
and since their paper several more papers have been written 
\cite{abroe_etal03,inst03}. 

\begin{figure}[t]
\centerline{\epsfig{file=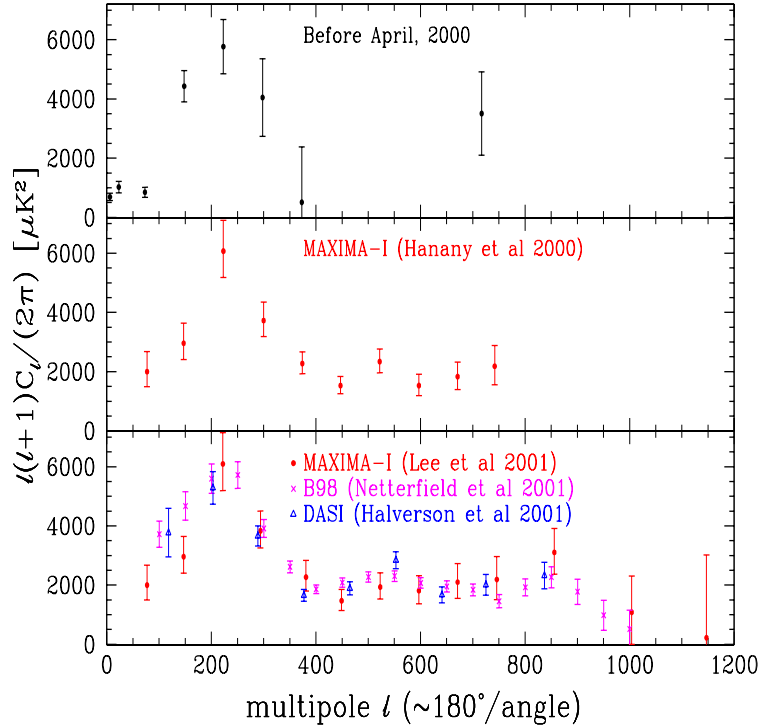,height=4.0in,width=4.0in,angle=-90}}
\caption{A combination of {\it all} CMB data prior to the first 
release of the MAXIMA and \boom\ data (top panel, courtesy of A. Jaffe),  
the \maxima\ data alone in 2000 \cite{hanany_etal00}, and 
the 2001 data of \maxima, \boom, and DASI 
\cite{boom_02,halverson_etal02,lee_etal01}. No calibration
adjustments have been made to the power spectra.}
\label{fig:pre_maxiboom}
\end{figure}

The \maxima-1 results had significant impact on cosmology. 
Together with the results from \boom\ \cite{boom_00}, they showed
conclusively that the geometry of the universe is close to flat
\cite{hanany_etal00}, and supported the evidence of 
\boom\ \cite{boom_02} and DASI
\cite{halverson_etal02} for harmonic peaks in
the power spectrum \cite{lee_etal01}.  Figure~\ref{fig:pre_maxiboom}
illustrates this leap in information content. 
The top panel in the figure, which is a combination 
of {\it all} the CMB data prior to April 2000, shows that indications that the
universe is flat were already evident in data of earlier experiments.
The middle panel shows {\it only} the 
\maxima\ data as released in May of 2000 shortly after
the release of the \boom\ data.  At the time of that first release the
\maxima-1 data gave the highest resolution map of the CMB and had
provided information over the broadest range of angular scales
compared to any other experiment.  To date MAXIMA has the highest
reported sensitivity of any CMB photometer and the highest combined
sensitivity of any CMB receiver\footnote{'Receiver sensitivity' is
defined as $\left[\Sigma_{i}
\left(1/\sigma_{i}^{2}\right)\right]^{-1/2}$ where $\sigma_{i}$ is
photometer sensitivity and the sum is over photometers from which
combined data is published.}.  Hanany et al.~\cite{hanany_etal00}
report photometer sensitivities as low as 80 $\mu K\sqrt{s}$ and a
combined sensitivity of 46 $\mu K\sqrt{s}$ for the \maxima-1 data set.
 
Jaffe et al. \cite{jaffe_etal01}
analyzed the accuracy with which the COBE-DMR, \maxima\ and \boom\
data constrain cosmological parameters when the datasets were analyzed
separately and together.  They found that the combination of COBE-DMR
and \maxima\ data constrained both the flatness of the universe and the
spectral index of the power spectrum of spatial fluctuations $n$ to 
within 9\% error (at $1 \sigma$).  The inclusion of the
\boom\ data improved the determination to within 6 and 9\%, 
respectively. The combination with other astrophysical data
showed that the universe is dominated by dark matter and 
energy~\cite{balbi_etal00,jaffe_etal01} .  
A year later, \maxima\, and \boom\ simultaneously 
released more of their data and DASI released new results.  The power
spectrum results of \maxima\ essentially have not changed, but were
extended to higher $\ell$ values. This 2001
collection of the data is shown in the bottom panel of
Figure~\ref{fig:pre_maxiboom}. 
Where they overlapped, all three power spectra were remarkably 
consistent with each other. DASI and \boom\ gave 
higher signal-to-noise ratio on the harmonic 
acoustic peak structure while \maxima\ had a broader coverage in $\ell$. 

The impact of all of these data was
that within a span of one year cosmology radically changed. 
Inflation gained strong supporting evidence, the framework of 
a universe overwhelmingly dominated by unknown forms of 
dark matter and energy had been transformed from a debated 
possibility to an essentially accepted fact, and the precision of the determination
of the cosmological parameters 
ushered what had been called the 'era of precision cosmology'. 
Subsequent data from other experiments and recently from
WMAP have confirmed these conclusions and significantly improved the
accuracy of the determination of all the cosmological parameters.

Before its release the \maxima-1 data were subjected to a battery
of systematic tests to ensure its validity.  The availability of data
from several independent photometers as well as the 
high redundancy of the scan strategy
provided multiple ways to cross-check the results and to ensure that
the contribution of systematic errors was negligible. In
Section~\ref{sec:m1_sys} of this paper we present the results of many
of these tests for the first time.

An even stronger systematic test is to cross-check the results against
those from an independent experiment.  We chose the scan region 
of \maxima-2 to partially overlap that of \maxima-1 
to allow a detailed comparison. In
Section~\ref{sec:maxima2} we give details of the \maxima-2 flight and
data analysis and present some comparisons between the \maxima-1 and
-2 data sets.  A recent, more detailed analysis has shown conclusively
that the \maxima-1, -2 and WMAP maps have detected the same spatial
fluctuations in a common region of the sky \cite{abroe_etal03}.

\section{Systematic Tests of the MAXIMA-1 Data}
\label{sec:m1_sys}

The \maxima\ instrument was reviewed in detail elsewhere 
\cite{inst03,hanany_etal00,lee_etalrome}. 
The \maxima-1 map, power spectrum
\cite{hanany_etal00,lee_etal01}, and cosmological results
\cite{balbi_etal00,stompor_etal01}
are based on the analysis of the combination of data collected by the
four photometers (three photometers for the Lee et al. paper \cite{lee_etal01})
that had the lowest noise equivalent temperatures (NET)
\cite{hanany_etal00}; hereafter we refer to them as $b34$, $b25$, $b45$ and
$b33$, where $b$ stands for 'bolometer' and the two digits define the
position of the bolometer in the $4 \times 4$ array.  The first three
detectors ($b34$, $b45$ and $b25$) operated at a frequency band
centered on $150$ GHz, and the forth ($b33$) at a frequency band
centered on $240$ GHz (data from $b33$ was not included in the 
results of Lee et al.)

We will discuss the following subset of systematics tests that 
have been carried out on the data: 
\begin{itemize}
\item a comparison of the maps and power spectra that were 
calculated from the data of individual photometers
(Section~\ref{sec:ind_phot}),
\item a comparison of maps and power spectra of a given 
region of the sky, but for which the data was taken 
at different times during the flight (Section~\ref{sec:temporal_comparison}), 
\item a comparison of the power spectra of different regions
of the map (Section~\ref{sec:spatial_comparison}). 
\end{itemize}
We will also discuss the effects of pixelization and noise
as they relate to the extraction of high $\ell$ information 
from the data (Section~\ref{sec:highell}).

All the maps presented in this section were
computed using a variation of the optimal maximum likelihood map-making
using the circulant noise approach
\cite{tegmark97b,stompor_etal02}. The maps have been pixelized
using square pixels of 8 arcminutes on a side and unless otherwise
noted are made from the data of all four photometers. When estimating
a power spectrum we deconvolved a circular top-hat pixel with an area
equal to that of the pixel \cite{wu_etal01a}.  'Sum maps' are
noise-weighted co-addition of constituent maps, and 'difference maps'
are half of the unweighted difference of the pixels common to both
maps.  Power spectra were computed using the quadratic estimator
approach \cite{tegmark97a,bjk98} with the MADCAP implementation
\cite{borrill99b} and are presented with bins in spherical harmonic
number $\ell$ of width $\Delta \ell = 75$.  The spectral bin
amplitudes have been decorrelated \cite{bjk98}. The theoretical 
power spectrum shown for reference in dotted line in some of the 
figures is the best fit cosmological model to the \maxima-1 data as given 
by Stompor et al. \cite{stompor_etal01}.

\subsection{Data of Individual Photometers}
\label{sec:ind_phot}

\subsubsection{Maps}
\label{sec:comp_maps}

Of the four photometers used for CMB data, the noise level of $b34$
was the lowest, achieving an NET of 80 $\mu K\sqrt{s}$ for most of the
flight, and $b33$ had the highest NET of 120 $\mu K\sqrt{s}$.  It is
therefore interesting to compare the maps and power spectra derived from
the data of these photometers.  In either case the time domain noise
properties were almost stationary throughout the entire CMB
measurement, not exceeding an end-to-end change in the white noise
level of 10-20\% in the most extreme cases.
\begin{figure}
\centerline{\epsfig{file=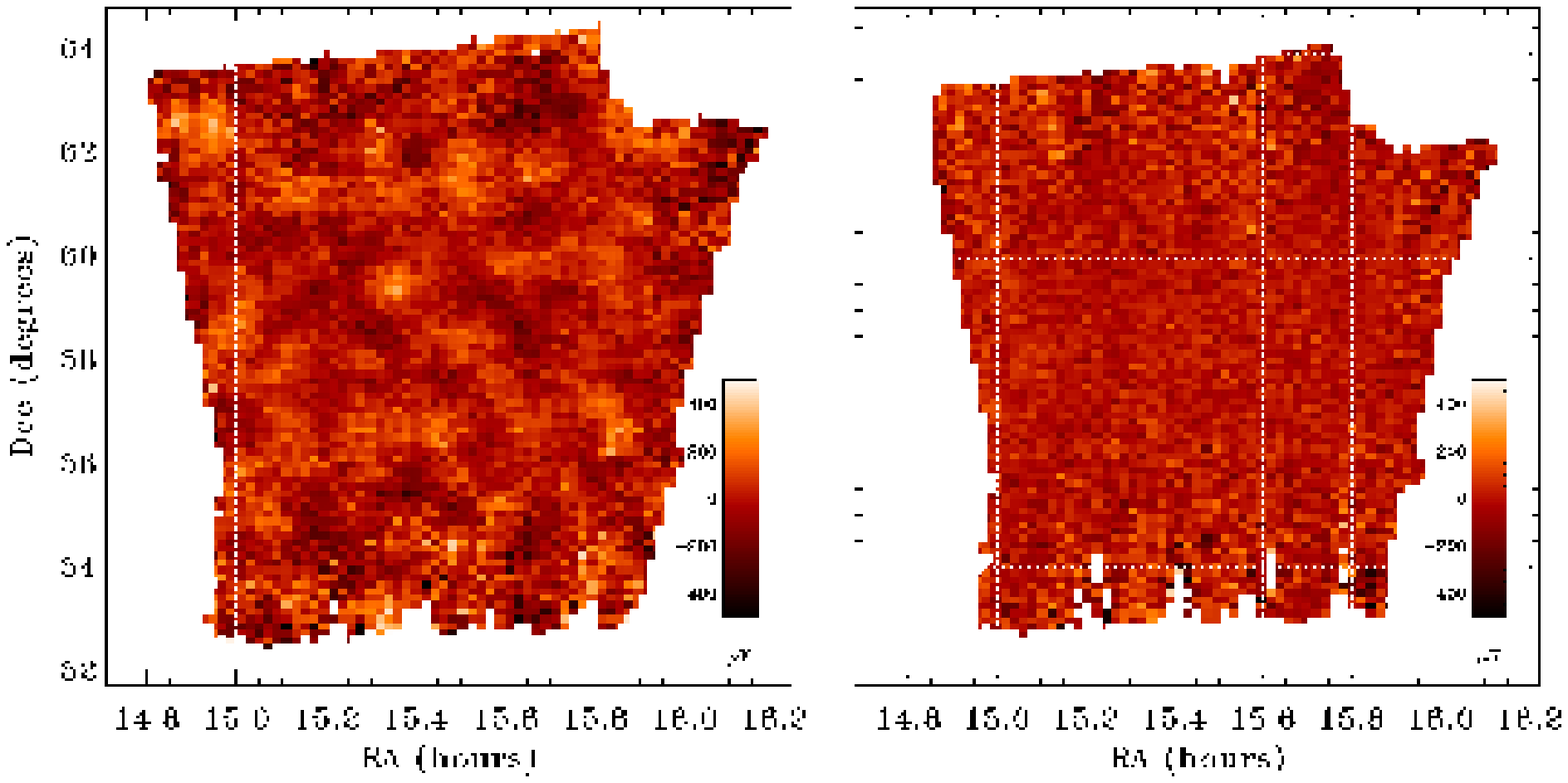,width=3.0in,angle=0}
\epsfig{file=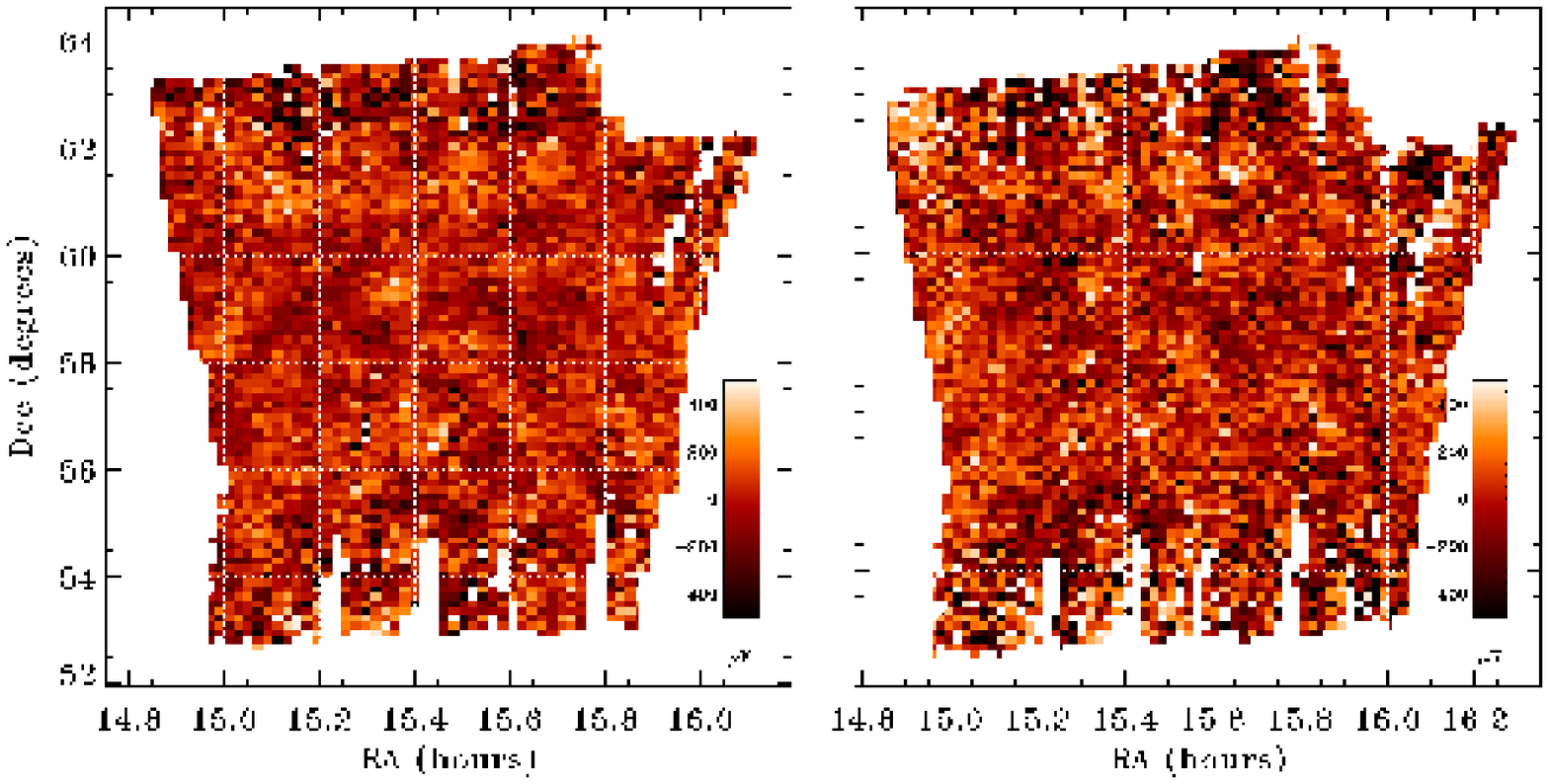,width=2.96in,angle=0}}
\caption{Left to right: the final map made from the data from all four
detectors combined, the map of the pixel noise
computed from the difference of two maps where each is made
from the combination of data from only two
detectors (i.e.\, $b34 + b45$ and $b25 + b33$), map from 
the data of $b34$ only, and map from the data of $b33$ only. 
Similar spatial fluctuations are present 
in all of the three maps that have a 
CMB signal and are absent in the map of the noise. }
\label{fig:ind_bolos}
\end{figure}
Maps made from the data of $b34$ \&\ $b33$ are shown in 
Figure~\ref{fig:ind_bolos} and show similar structure
throughout the map but more predominantly in the low-noise central
part of the maps. The same sky structure is also readily discernible
in the map made by combining the data of all four detectors, but the
structure disappears in the four-detector $((b34+b33)-(b45+b25))$
difference map.  This visual impression is expressed
quantitatively using the following statistics: \\
$\bullet$ the $\chi^2$ statistic,
\begin{equation}
\chi^2 \left(\bf{m}\right)  \equiv  \bf{m}^T {\cal N}^{-1} \bf{m}, 
\,\,\,\,\,\,\,\,\,\,\,\,
\kappa \left(\bf{m}\right)  \equiv  {\left(\chi^2\left(\bf{m}\right) -
  n_{\rm DOF}\right) \over \sqrt{2 n_{\rm DOF}} }
\end{equation}
where $\bf{m}$ and ${\cal N}$ denote a map and a pixel-pixel noise
correlation matrix respectively. The statistic has $n_{\rm DOF}$
effective degrees of freedom, which are assumed to be equal to the
difference between the number of pixels and the low
$\ell$-modes that are removed from the
map prior to the analysis.  
Assuming Gaussian noise, the coefficient $\kappa$ quantifies the
distance in units of standard deviation between the computed value of
$\chi^2$ and the value expected if no sky signal was present in the
map; \\
$\bullet$ the ``null buster'' statistic $\nu$ \cite{tegmark99},
\begin{equation}
\nu \left(\bf{m}\right)\equiv {\bf{m}^T {\cal N}^{-1}{\cal S}{\cal N}^{-1}\bf{m}-{\rm Tr}\left[ {\cal N}^{-1}{\cal S} \right]
\over \left\{2 {\rm Tr}\left[{\cal N}^{-1}{\cal S}{\cal N}^{-1}{\cal S}\right]\right\}^{1/2}},
\end{equation}
where ${\cal S}$ is an arbitrary matrix, which is equal to the signal
correlation matrix computed for the best fit \maxima-1 power spectrum
smoothed with a Gaussian beam of 10 arcminutes full-width at half
maximum and an axially symmetric (approximate) pixel window function
\cite{tegmark99}.  This statistic determines the number of standard
deviations at which a given map $\bf{m}$ is inconsistent with 
a hypothesis of only having noise, given that the signal is described by the
correlation matrix ${\cal S}$ (and it is therefore analogous to the
$\kappa$ statistic, with which it coincides if ${\cal S}={\cal N}$); \\
$\bullet$ the one-dimensional Kolmogorov-Smirnov (KS) test 
applied to noise-prewhitened maps \cite{stompor_etal02} and defined as,
\begin{equation}
\widehat{\bf{m}}\equiv {\cal F}^{-1/2} \bf{m},\,\,\, \mbox{where ${\cal F}$ is
assumed to be a symmetric matrix such as}\,\,\,\, 
{\cal N}\equiv {\cal F}^{1/2}{\cal F}^{1/2}. 
\end{equation}
For each map we compute a KS significance coefficient
giving the confidence level at which the hypothesis that the prewhitened
map has been randomly drawn from the Gaussian distribution with a unit
variance can be accepted; \\ 
$\bullet$ the probability enhancement factor $\beta$ \cite{knox_etal98},
\begin{equation}
\beta\left(\bf{m}_i,\bf{m}_j\right) \equiv \ln \left\{ 
\frac{ {\cal P} \left( \left[
\begin{array}{c}\displaystyle{\bf{m}_i}\\\displaystyle{\bf{m}_j}\end{array}
\right] | \left[
\begin{array}{c c}\displaystyle{{\cal C}_i} & \displaystyle{{\cal C}_{ij}} \\
\displaystyle{{\cal C}_{ji}} & \displaystyle{{\cal C}_j} \end{array} 
  \right] \right)} {\displaystyle{
{\cal P} \left( \bf{m}_i \vert\;{\cal C}_i \right) 
{\cal P} \left( \bf{m}_j|\;{\cal C}_j\right) }} \right\}
\end{equation}
where the matrix ${\cal C}_i$ describes the CMB signal correlation
matrix computed for a map $\bf{m}_i$, and ${\cal C}_{ij}$ is the
signal cross-correlation matrix for maps $\bf{m}_i$ and ${\bf
m_j}$. In our case both are computed assuming the \maxima-1 best fit
power spectrum smoothed with the antenna beam and the pixel window
function for the null-buster statistic.  The quantity ${\cal
P}\left({\bf m_i} |\; {\cal C}_i\right)$ represents the probability
distribution of realizations of maps with signal correlations given by
${\cal C}_i$ and noise correlations given by ${\cal N}_i$; we 
assume that the probability distribution is 
a multi-variate Gaussian.  We assign a statistical
significance to this statistic by computing its mean and variance
either under the assumption of no correlation or the assumption of
full correlation of the sky signal in both maps
\cite{knox_etal98,coble_etal01}.  We denote these value
$\beta_0\left(\bf{m}_i,\bf{m}_j\right)$ and
$\beta_\infty\left(\bf{m}_i,\bf{m}_j\right)$ respectively.
\begin{table}[t]
\begin{center}
\begin{tabular}{c c c c c c c c}
\hline 
$\bf{m}_1$ & $\bf{m}_2$ & $\kappa $ & $\rm{KS} $ & $\nu $ & $\beta $ & $\beta_0$ &
   $\beta_\infty$ \\
\hline
b34     & b45     & -1.9  & 28\% & -0.4 & -180 & $-166\pm14$ & $-2537\pm201$ \\
b34     & b25     & -0.06 & 44\% & -0.8 & -198 & $-184\pm15$ & $-2557\pm212$ \\
b34     & b33     & -0.2  & 49\% & -0.5 & -146 & $-142\pm14$ & $-2644\pm439$ \\
b45     & b25     & -0.7  & 91\% & -0.3 & -172 & $-171\pm15$ & $-2444\pm106$ \\
b33     & b45     &  0.1  & 22\% &  0.6 & -128 & $-129\pm13$ & $-2454\pm121$ \\
b25     & b33     & -0.02 & 84\% & -0.9 & -143 & $-139\pm13$ & $-2464\pm127$ \\
b34+b33 & b45+b25 &  0.3  & 90\% & -0.6 & -292 & $-281\pm18$ & $-2973\pm234$ \\
CMB1    & CMB2    & -1.6  & 81\% &  0.1 & -168 & $-195\pm15$ & $-1858\pm132$ \\
\hline 
\end{tabular}
\caption{Results of statistical tests that were applied to the differences
of pairs of maps (columns 3, 4 and 5) that were produced from the
data of photometers listed in columns 1 and 2, 
the probability enhancement factor
test (column 6) applied to the same pairs of maps, and
the expected average and $68\%$ (``$1\sigma$'') confidence ranges
under the hypothesis of perfect (column 7) or lack of (column 8)
correlations of the signal in both maps. Columns 3, 4 and 5 have
results for the $\chi^2$, Kolmogorov-Smirnov and null-buster
statistics, respectively, and show very good consistency with the
hypothesis that the difference maps contain no sky signal. }
\label{tab:glob_stat}
\end{center}
\end{table}

Due to the small size of the \maxima-1 maps the very low-$\ell$
content of the maps may not be reliable so one may not want to include
it in the tests described here. Therefore in the case of the $\chi^2$,
``null-buster'' and KS statistics we ``weighted out'' \cite{bjk98} 
all the $\ell$-modes with $\ell \le 35$ by replacing
the inverse noise correlation matrices ${\cal N}^{-1}$ by
\begin{equation}
{\cal N}^{-1} \longrightarrow {\cal N}^{-1} - 
  \left({\cal N}^{-1}{\cal B}\right)^T\left[
{\cal B}^T{\cal N}^{-1}{\cal B}\right]^{-1}
\left({\cal N}^{-1}{\cal B}\right), \,\,\,\,\mbox{where} \,\,\,\,\, 
{\cal B}_{ik}\equiv \sum_{j}\psi^i_{j} \left[\psi_j^{k}\right]^T
\end{equation}
and the $\psi$ constitute a set of linearly independent pixel vectors
spanning the same space as all the spherical harmonics with $\ell \le
35$; $\psi^{i}_{j}$ is the $i$-th pixel
component of the vector $\psi_{j}$.
This correction corresponds to assigning ``infinite'' noise to the
spatial modes described by the functions $\psi$
\cite{bjk98,stompor_etal02}.  Consequently these modes do not
contribute to final results of any of the statistics.  For the
probability enhancement factor we have applied an analogous correction
to the inverse (signal+noise) correlation matrix, ${\cal S}+{\cal N}$.
For the sky patches considered here we usually find that there are
only $\simeq 55$ independent modes (and hence vectors $\psi$) out of a
total of $1296$ spherical harmonics with $\ell \le 35$.  We have also
found that although the particular values of the statistical tests
depend on whether the modes with $\ell \le 35$ are rejected or not,
the overall conclusions remain essentially unchanged.
\begin{table}[t]
\begin{center}
\begin{tabular}{c c c c c c c c c c}
\hline 
   & b34 & b45 & b25 & b33 & B34+b45 & b25+b33 & CMB1 & CMB2 & ALL \\ \hline
$\kappa\left(\bf{m}\right)$ & 16 & 11 & 20 & 7 & 30 & 28 & 33 & 24 & 69 \\ 
$\rm{KS}\left(\bf{m}\right)$ & 0\% & 0.5\% & 0\% & 1\% & 0\% & 0\% & 0\% &0\%&0\% \\
$\nu\left(\bf{m}\right)$ & 70 & 60 & 100 & 35 & 130 & 135 & 140 & 140 & 317 \\
\hline
\hline
\\
\end{tabular}
\caption{Results of the same statistical tests as shown in
Table~1 but applied to single photometer maps. Since these maps do
contain sky signal, contrasting these results with those in Table~1
demonstrates the sensitivity of each statistic to the presence of sky
signal in the map. Zeros in the case of the KS statistics stand for
numbers less than $10^{-7}.$ }
\label{tab:sing_phot}
\end{center}
\end{table}

The results of these tests as applied to various pairs of maps are
given in Table~\ref{tab:glob_stat}.  They
confirm the visual agreement between the maps that were produced from
the data of different detectors.  The absolute values of $\kappa$ and
$\nu$ computed for the difference maps are usually $\simlt 1$, and
always less then $2$, which is to be
interpreted as a ``better than $2 \sigma$'' agreement.  For the
probability enhancement factor, the value of $\beta$ always agrees
with the expected value of $\beta_0$ within the quoted ``$1\sigma$''
uncertainty and always disagrees by more than ``6 $\sigma$'' (and
usually $\sim 15-20 \sigma$) with the appropriate $\beta_\infty$.
The latter values are expected for $\beta$
if there is no correlation between a given pair of maps. Note that both the
null-buster statistic and the probability enhancement factor depend on
the choice of the signal power spectrum.  However we have found that
if we adopt a flat power spectrum rather than the best-fit spectrum
chosen above then the numbers computed for these statistics change by
no more than 10-20\% and their statistical interpretation remains the
same \cite{tegmark99}.

We have also applied the $\chi^2$, null-buster and KS statistics to
the single detector maps.  The results are collected in
Table~\ref{tab:sing_phot} and show that a strong signal is detected 
in all cases.   In the case of the null-buster
test, the numbers computed here can be compared with those obtained
for the Saskatoon and QMAP experiments which are $21$ and $40$,
respectively \cite{xu_etal02}. It is clear that according to this
statistic there is more 
information content in a map made from a single 
detector of \maxima-1 than there is in the final maps produced by
either of those experiments.  When contrasted with the values obtained
for the difference maps, these numbers can be viewed as a
demonstration of the sensitivity of the tests. However it is important
to bear in mind that the noise level and correlations are different in
the two-detector difference maps than in any single detector map.

\subsubsection{Power spectra}

\begin{figure}
\centerline{\epsfig{file=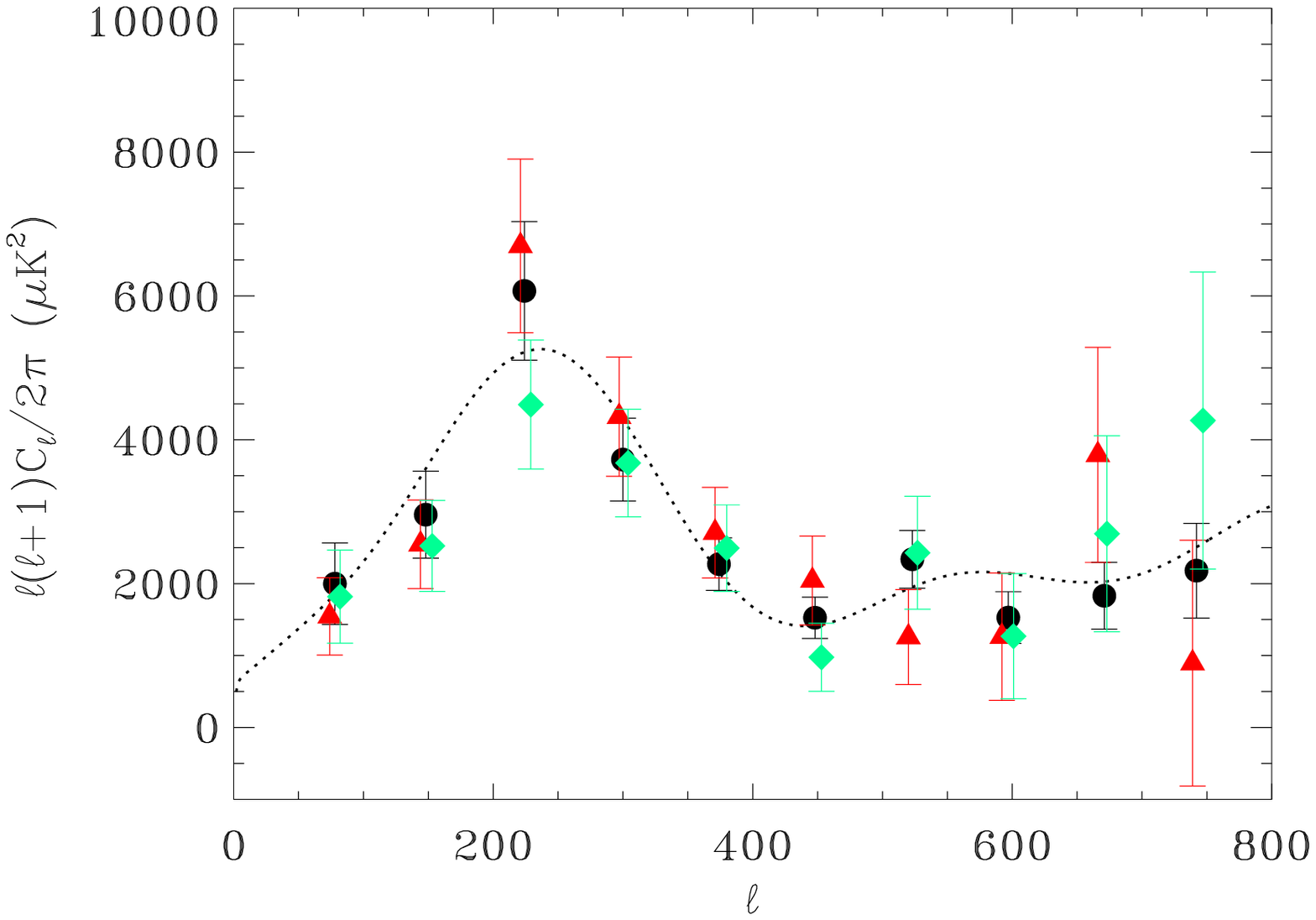,width=3.0in,angle=0}
\epsfig{file=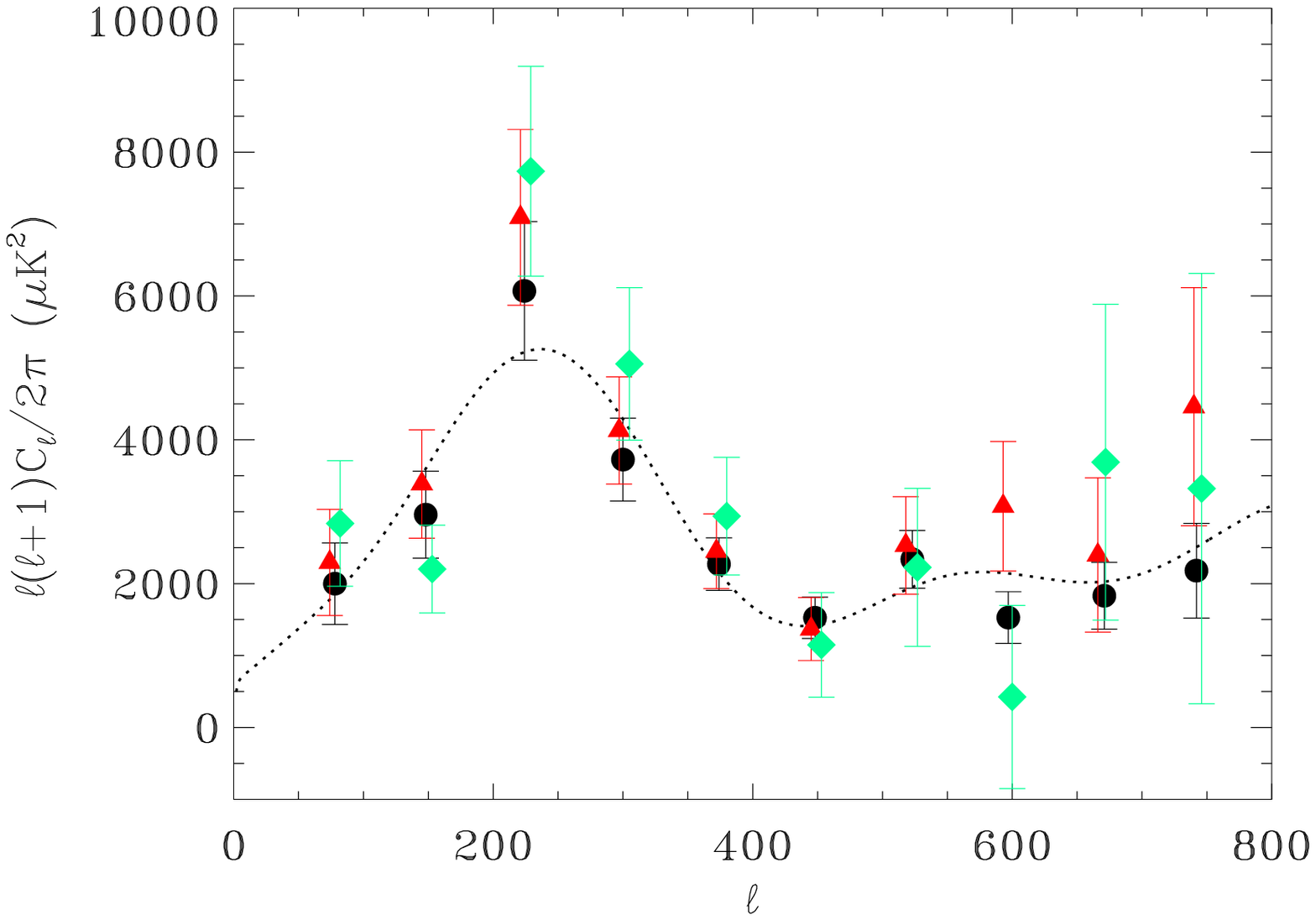,width=3.0in,angle=0}}
\caption{Angular power spectra using the data of 
$b34$ ($b25$) in triangles and of $b45$ ($b33$) in diamonds
in the left (right) panel. The power spectrum from the 
combined data from all four detectors (circles) \cite{hanany_etal00}
is also shown for comparison. 
In each bin the
triangles and diamonds have been displaced slightly from the true
central values (shown by the filled circles) to make the figure
readable. The dotted curves is a best fit cosmology to the 
MAXIMA-1 results \cite{stompor_etal01}. }
\label{fig:single_chan_spectra}
\end{figure}

The power spectra for each of the detectors individually and combined
(Figure~\ref{fig:single_chan_spectra}) are consistent throughout the
entire $\ell$ range, with the scatter in the estimated bin power
increasing at the higher and noisier $\ell$ bins. The error bars
plotted here reflect minimally correlated statistical uncertainty
only, and exclude any fully correlated systematic uncertainties. Such
systematic uncertainties could come from an overall misestimation of
the calibration, which has the effect of renormalizing the entire
power spectrum, or from beam reconstruction uncertainty, which is
important predominantly at high $\ell$. The calibration
uncertainty is about 8\% in power for the data of any single
photometer and we have conservatively assumed a combined calibration
uncertainty of 8\% for the combination of all photometers.  The beam
reconstruction uncertainty of \maxima-1 has been investigated in 
great detail by Wu et al.~\cite{wu_etal01a} and Lee et
al.~\cite{lee_etal01}.

\begin{figure}
\vspace{-2cm}
\centerline{\epsfig{file=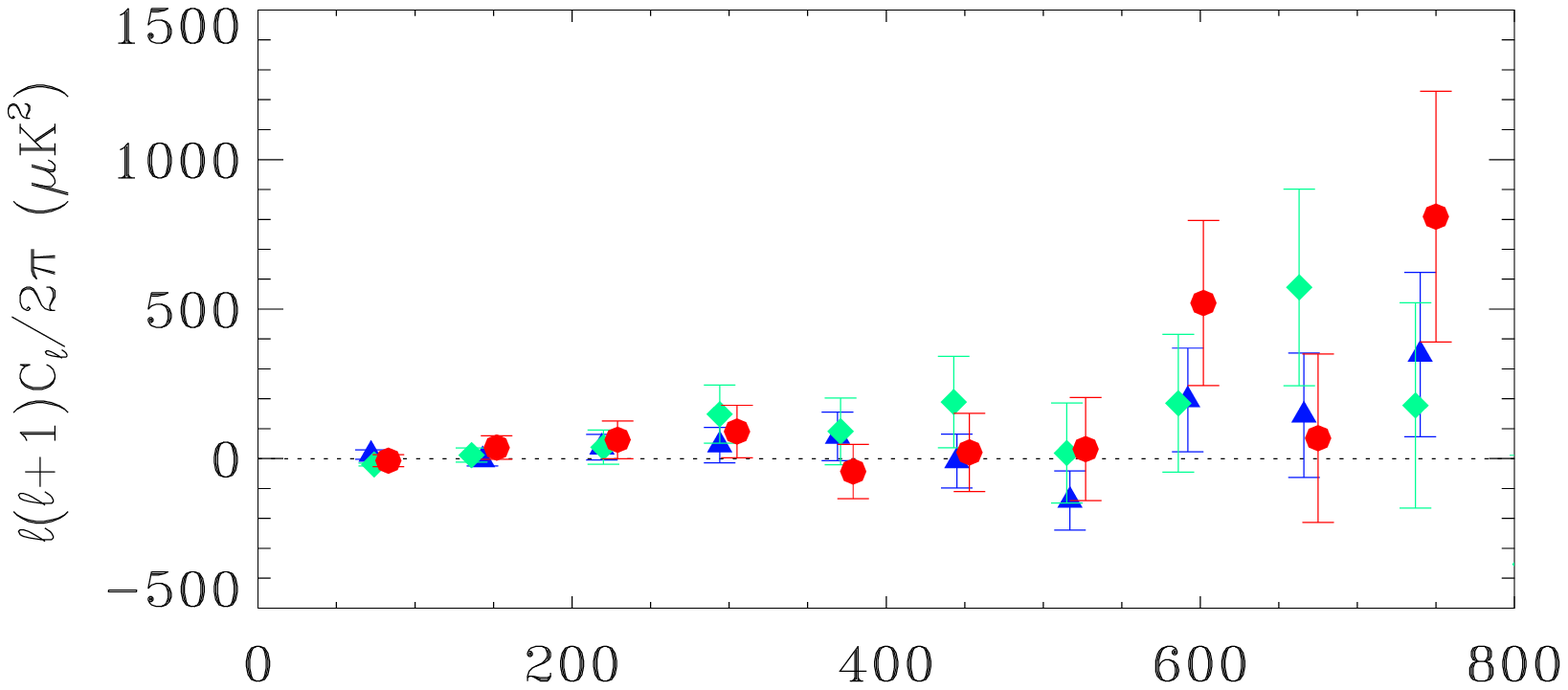,width=3.0in,angle=0}
\epsfig{file=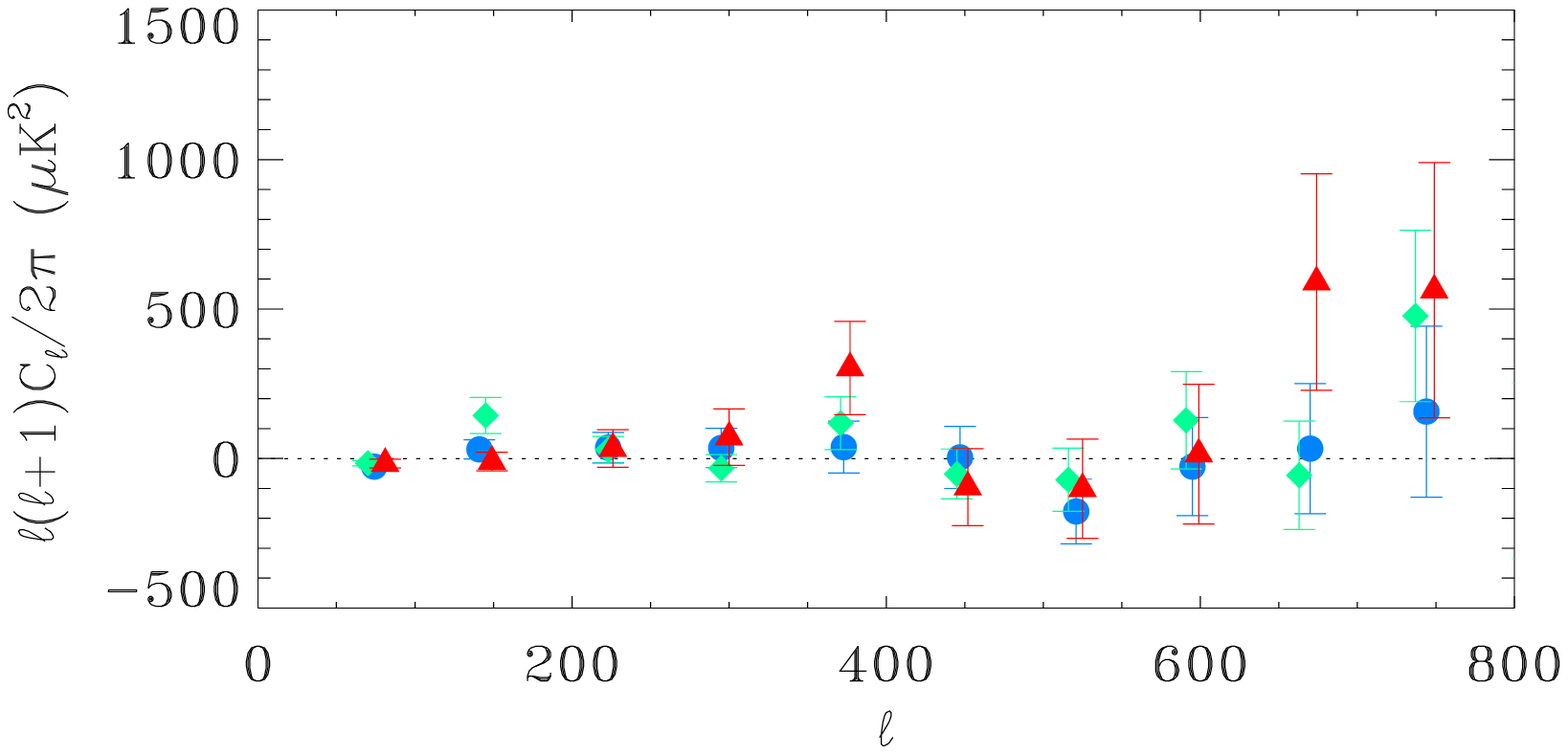,width=3.0in,angle=0}}
\caption{Angular power spectra
of the single detector difference maps. The left panel shows the
three combinations excluding, and the right panel including, the $b45$
photometer.}
\label{fig:single_chan_diffspectra}
\end{figure}

\subsubsection{Difference maps}

Differencing two maps of the same patch of the sky is a sensitive
method of searching for systematic problems in the data. Power spectra
of such difference maps -- unlike the ``single number'' statistics of
the Section~\ref{sec:comp_maps} -- may not only detect a problem but also locate
the angular scale at which it occurs, thereby providing a useful
diagnostic.

From the four single detector \maxima-1 maps we form six distinct,
although not independent, difference maps; the power spectra
of these difference maps are shown in
Figure~\ref{fig:single_chan_diffspectra}.  A $\chi^2$ with a null
model gives values of $\simeq 1$ per degree of freedom for all
differences.  The only points deviating from zero by more than
$2\sigma$ are found at the very low-$\ell$ end of the power
spectra. This is not surprising given the difficulty of estimating the
lowest frequency noise modes in the time domain \cite{stompor_etal02},
which dominate the noise contribution on large angular scales.  If we
interpret any residual power as an estimate of a systematic error, we
find that the magnitude of such an error is much smaller than the
statistical uncertainty in the power in the corresponding bins.

The slight excess of positive detections over negative at high $\ell$
(which may appear to be a trend, but in fact does not continue to yet
higher $\ell$, see for example the right panel 
of Figure~\ref{fig:three_vs_eight}) is most likely the
residual of sky signal that persists in the difference maps due to the
somewhat different beams of the various detectors. Although such a
signal is expected to be rather small, it is amplified by the
deconvolution of the beam and pixel window function in the power
spectrum estimation. The error budget of the final \maxima-1 
spectrum~\cite{hanany_etal00}
includes the effect of differences of beams between different 
detectors but the effect is not included when calculating  
difference spectra such as shown in Figure~\ref{fig:single_chan_diffspectra}. 


\subsection{Temporal Comparison}
\label{sec:temporal_comparison}

\begin{figure}[t] 
\centerline{\epsfig{file=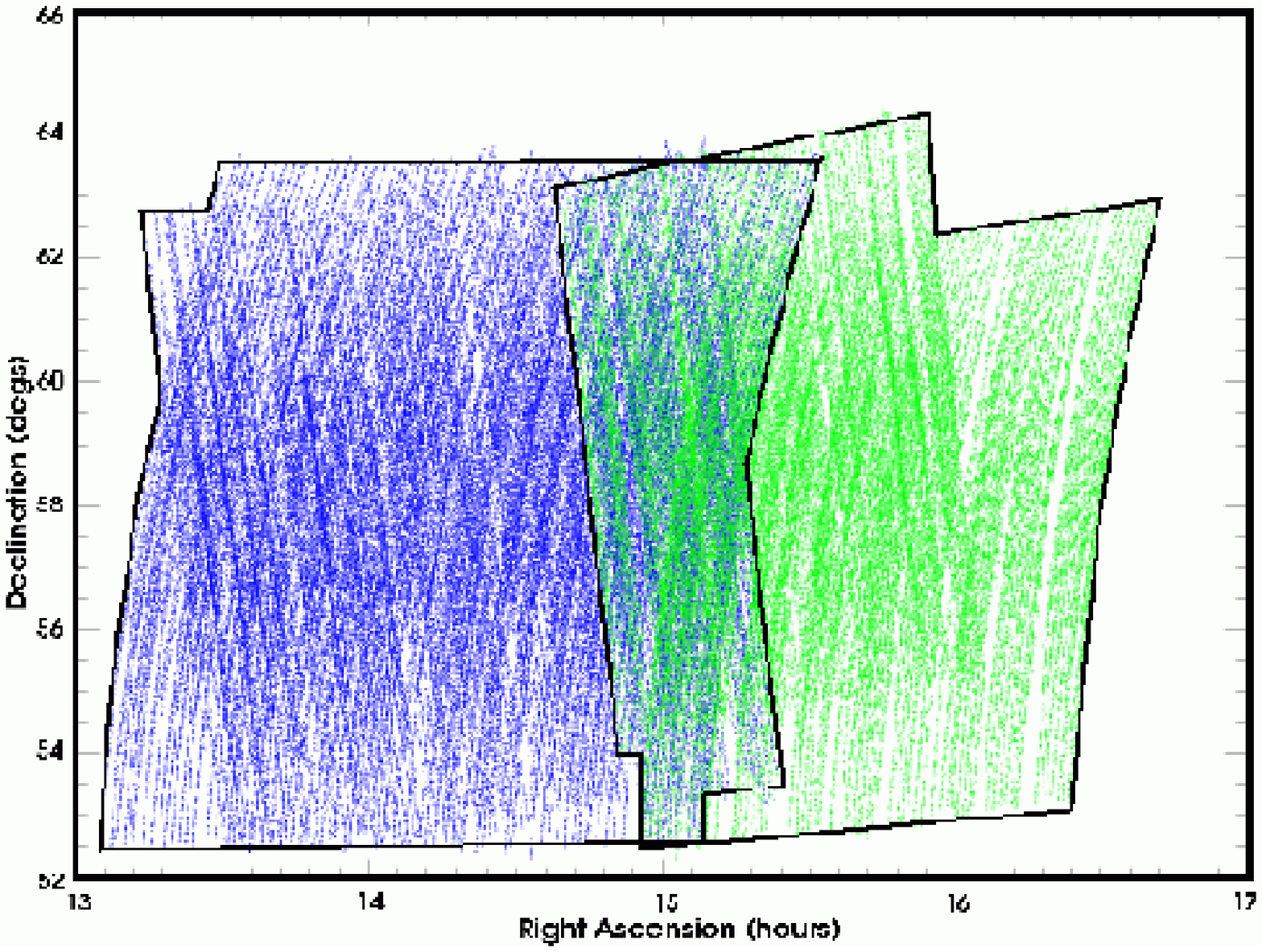,width=1.5in,height=1.5in,angle=0} \epsfig{file=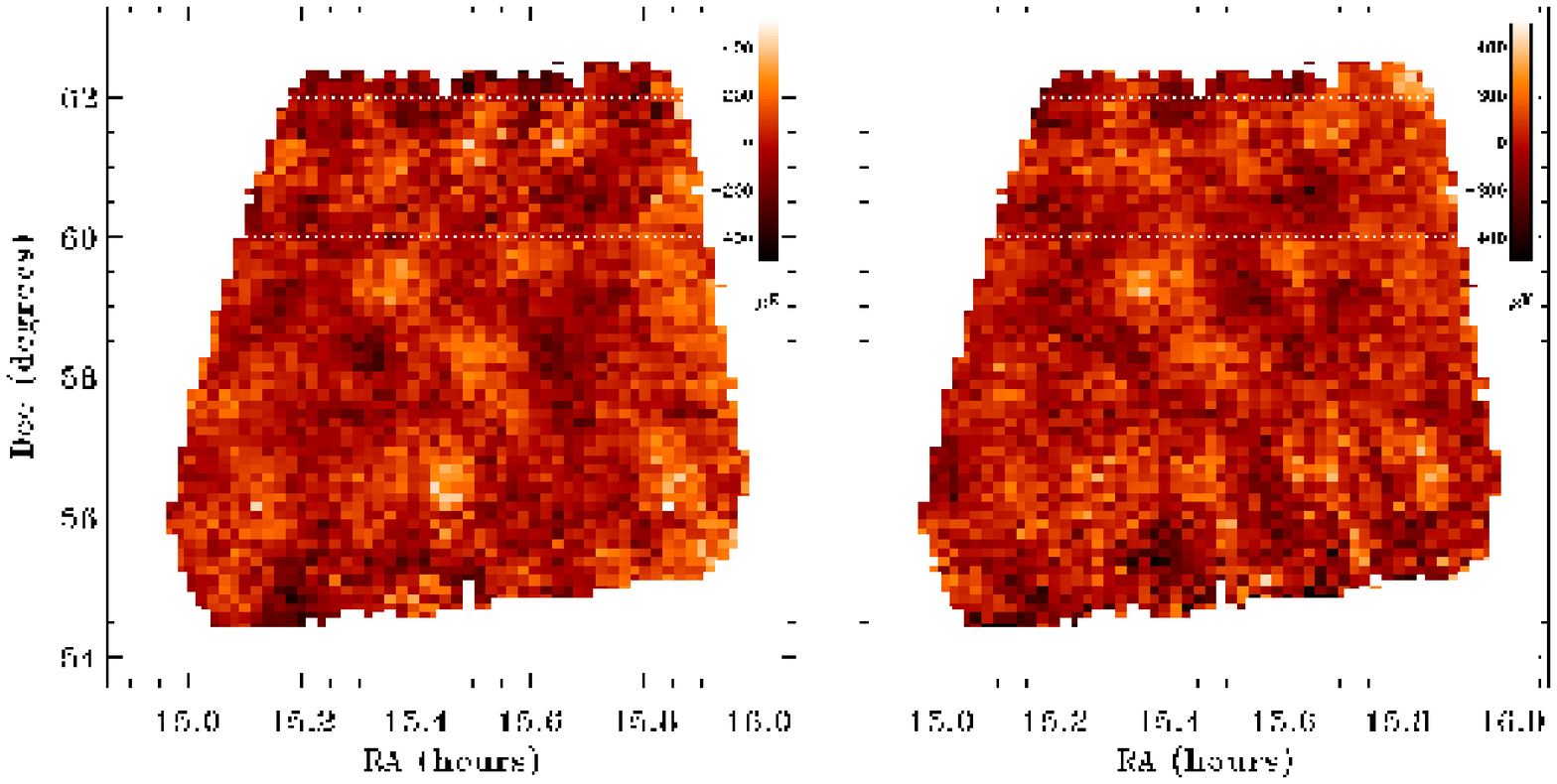,width=3.0in,angle=0}}
\caption{Left: the areas of the sky scanned by \maxima-1 (green) and 
\maxima-2 (blue). Each
of the areas was scanned with a 'CMB1' and 'CMB2' distinct scans that 
were taken at different times and that have a relative angle of about 
25 degrees. Each point in the plot represents a pointing of the telescope
averaged over $\sim 100$~msec.  
The \maxima-2 area overlaps about 50 square degrees of the area of \maxima-1 
providing an important systematic test. \newline
Right: Maps of the \maxima-1 CMB1 (left panel) and CMB2
(right) scans. Only the overlapping region of both scans is shown.}
\label{fig:max12scans}
\end{figure}

\begin{figure}[t]
\centerline{\epsfig{file=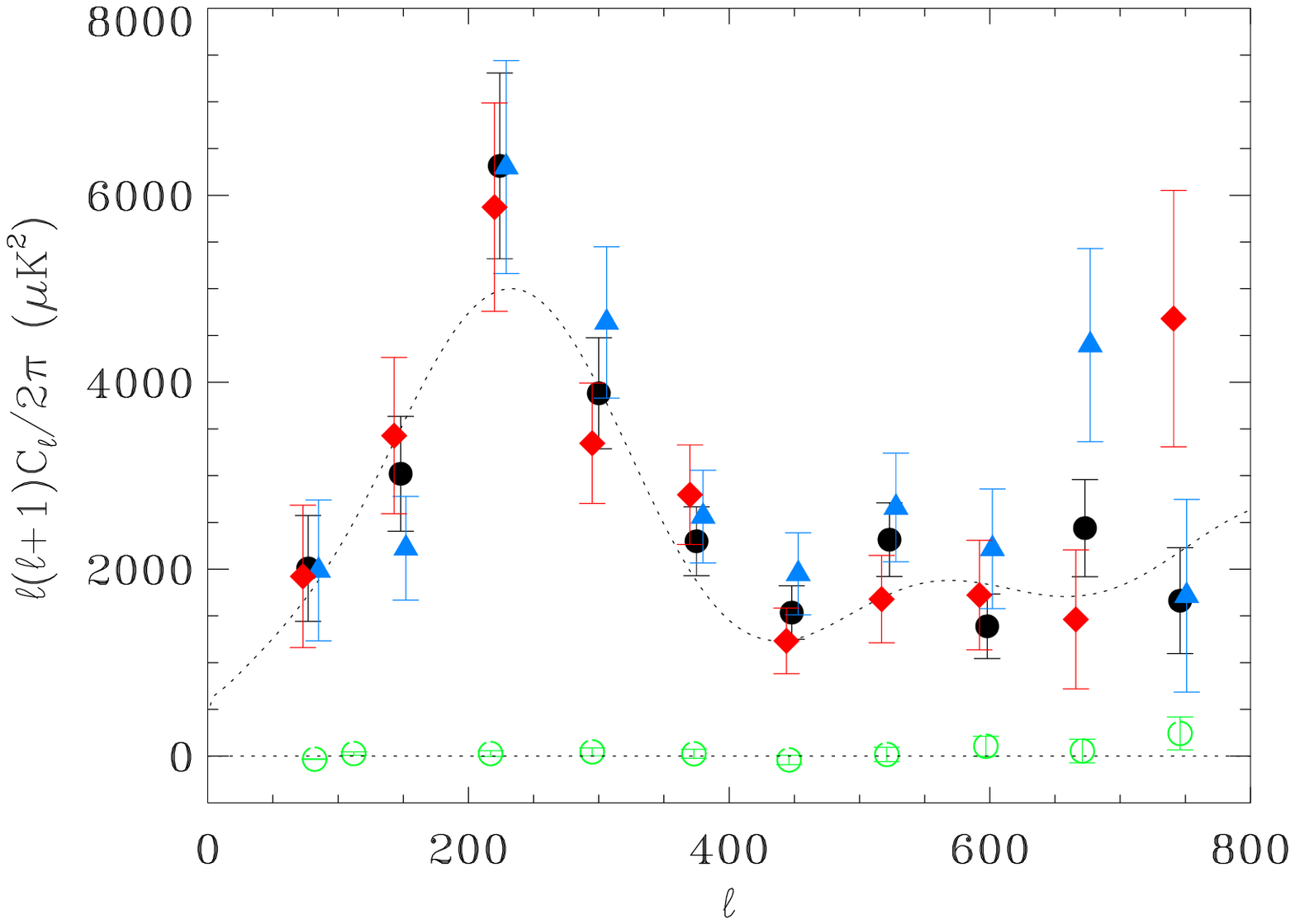,width=3.0in,angle=0}
   \epsfig{file=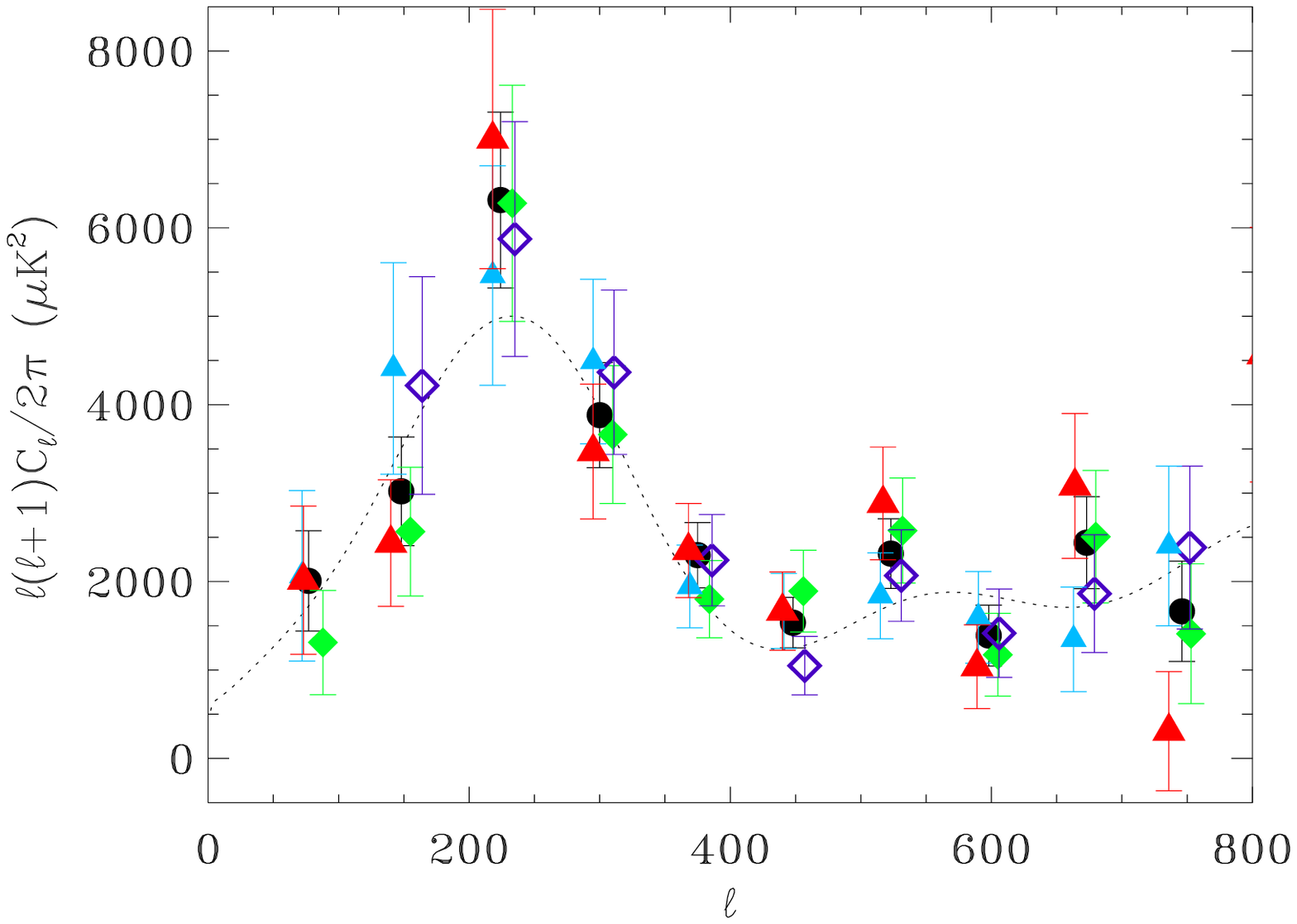,width=3.0in,angle=0} }
\caption{Left: angular power spectra
of the CMB1 (diamond) and CMB2 (triangle) scans using data from four detectors,
the spectrum published by Hanany et al.\ \cite{hanany_etal00}(filled circles) 
and spectrum of the difference map. Right: angular power spectra computed 
for the left and right
halves of the map (filled and open diamonds, respectively) and the
upper and lower halves (filled and open triangles) as well as the full
map (filled circles). Each of the sub-maps contains only $\sim 3000$
pixels.}
\label{fig:cmb12_powerspectra}
\end{figure}

During the 1998 flight of \maxima\ each photometer observed the same
patch of the sky twice, with an approximately 90 minute gap between
observations. This provides a natural division of the data into two
parts, which we call CMB1 and CMB2. These scans are the two green 
shaded areas in left panel of Figure~\ref{fig:max12scans}, which are oriented at
an angle of about 20 degrees to each other. Because of the time
lag between the scans the maps of CMB1 and CMB2 
may have different systematic errors and it is valuable to 
compare them. The maps made from the combination of the data 
from four photometers are shown in the two right panels of 
Figure~\ref{fig:max12scans}. 
We can clearly see that the structure is generally well
replicated in each map. The visual impression is confirmed when 
we calculate the statistics of Section~\ref{sec:comp_maps} to 
compare the maps, and also when we calculate the corresponding 
power spectra, which agree well both with one another and with the
``canonical'' \maxima-1 spectrum; see the left 
panel in Figure~\ref{fig:cmb12_powerspectra}.  
Some concern might be
raised by the bright spot in the CMB2 map (at RA$\simeq 15.7$ hours
and DEC$\simeq57$ degree) which has no counterpart in CMB1.  We do not
expect this feature to have any bearing on the final results, although
we have failed to single out an unambiguous source for the difference,
or even to determine its statistical significance. This
is probably an artifact of the map-making algorithm due to poor
cross-linking in this region. This suspicion is supported by the observation that no
feature of this sort is found in the better cross-linked map combining
the data from both scans. Moreover, applying the power spectrum
analysis to maps with the pixels corresponding to this feature removed
shows no significant change in the results.

\subsection{Spatial Comparison}
\label{sec:spatial_comparison}

An interesting test of the data is to compute and compare the power
spectra of sub-maps of the entire map. Such sub-map spectra should
agree to within the sampling and noise variances. The disadvantage of
this approach is that because of pixel-pixel noise and sky signal
correlations, the interpretation of differences between the spectra
obtained is not straightforward. Furthermore the uncertainties in the
sub-map spectra rapidly grow as the number of pixels decreases, making
comparisons between small sub-maps meaningless. Here we investigate
two halving subdivisions of the full map -- left versus right and top
versus bottom. These spectra are shown in the right panel of 
Fig.~\ref{fig:cmb12_powerspectra} and are in good agreement.

\subsection{The High $\ell$ Regime}
\label{sec:highell}

The first release of the \maxima-1 data \cite{hanany_etal00} included
information only up to $\ell=785$ because more time and computational 
effort was required to ensure 
that all systematic errors have been analyzed thoroughly for the higher $\ell$ 
regime. In the second release \cite{lee_etal01} a subset of the data 
from the first release
was analyzed to give information up to $\ell=1200$. 
Here we discuss how this subset of the data was chosen. 

\subsubsection{Spatial Cut}

Pixelization of the maps introduces an extra smoothing of the
underlying CMB signal on very small scales.  Applying an appropriate
window function to compensate for the smoothing (as described earlier)
assumes an unrealistic perfect sampling of every pixel in the map. In
reality the smoothing introduced by the pixelization procedure is
position (pixel) dependent, and difficult to deconvolve exactly from
the final spectrum.  One solution is to decrease the pixelization
scale until the smoothing that it induces does not affect the spectrum
in the range of $\ell$ of interest. However, this has to be weighed
against the increased computational cost of analyzing maps with more
pixels.  Another solution is to use relatively big pixels but include
only pixels that happen to be sampled very uniformly and for which the
smoothing should be well characterized by the approximate window
function.
\begin{figure} 
\centerline{\epsfig{file=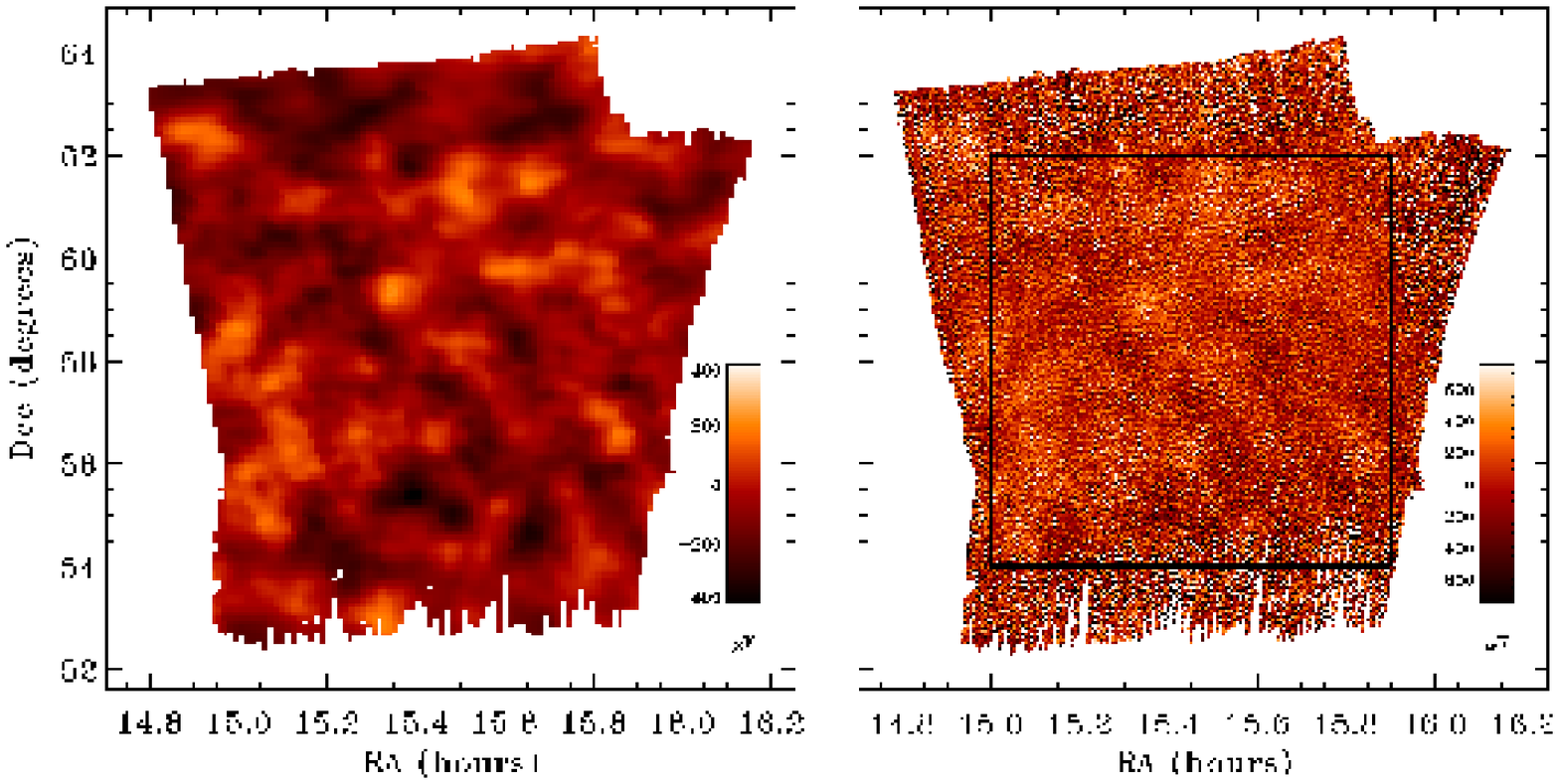,width=5.0in,angle=0}}
\caption{The entire \maxima-1 map pixelized with 3 arcminute pixels
and a demarcation of the region used to produce the high $\ell$ 
region of the power spectrum (right) and a Wiener filtered version 
of the map (left). The color stretches are $\pm 750$ $\mu$K and 
$\pm 400$ $\mu$K, for the right and left panels, respectively.}
\label{fig:3arcm_map}
\end{figure}

\begin{figure}
\centerline{\epsfig{file=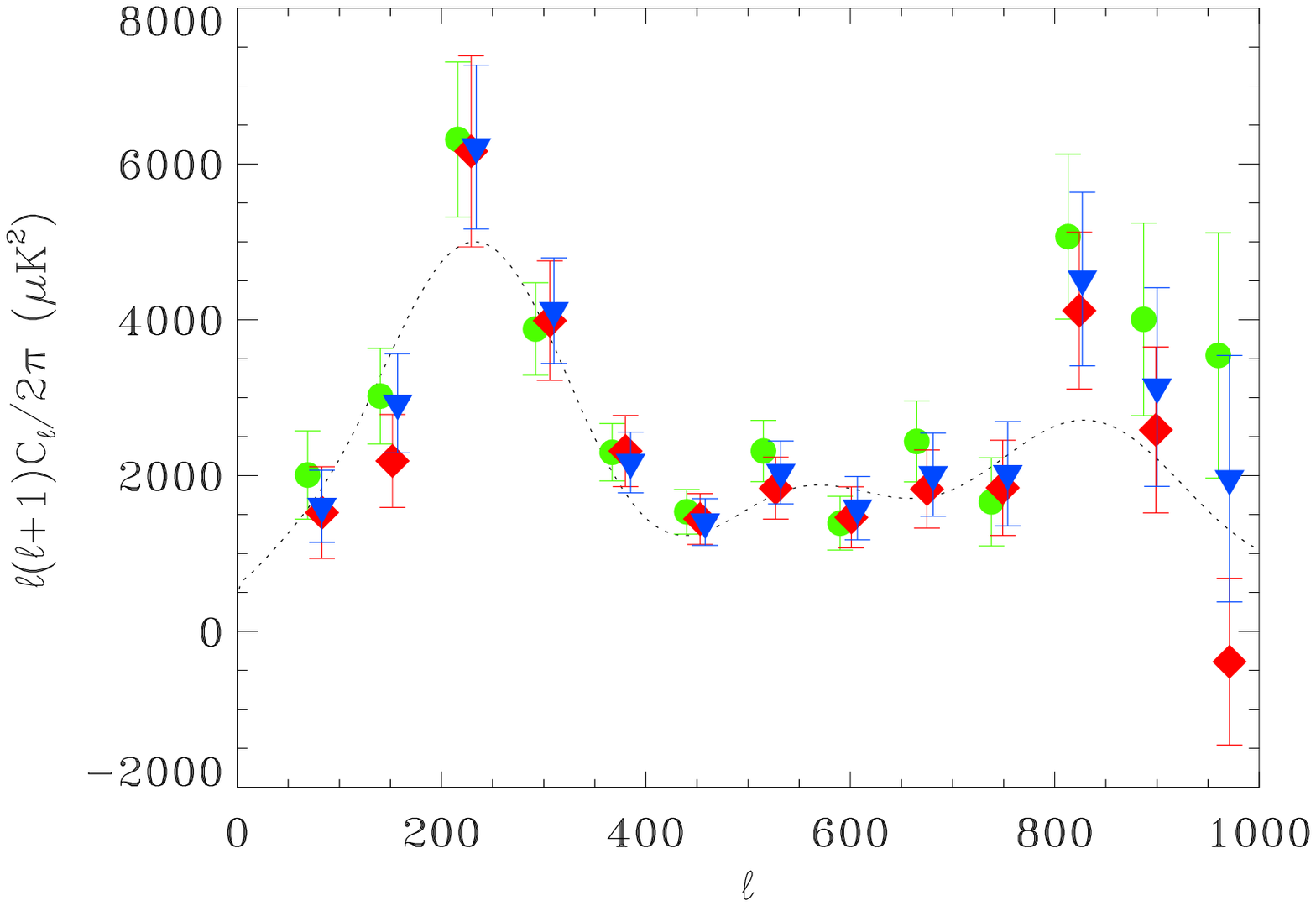,width=3.0in,angle=0} 
\epsfig{file=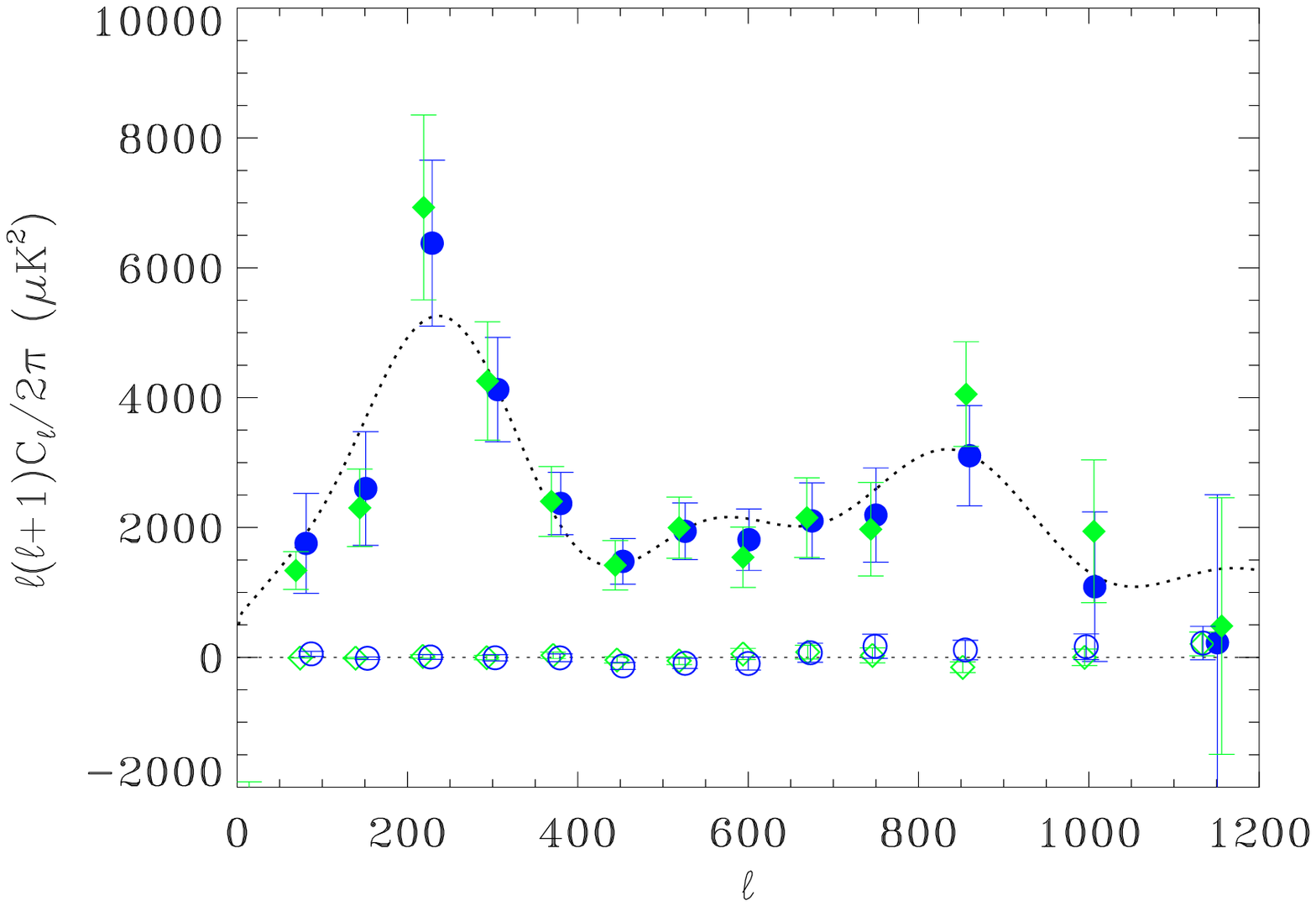,width=3.0in,angle=0} }
\caption{Left: angular power spectra
of a map with 8 arcminutes pixels (circles), only 
the central section of 
the map also pixelized with 8 arcminutes pixels (triangles) and 
the entire map with 3 arcminutes pixels, but with no deconvolution 
of a pixel window function (diamonds); see text. \newline
Right: power spectra of maps pixelized with 3 arcminutes 
and made using data from 
all four detectors (diamonds) and including only three
detectors by excluding the data from $b33$ (circles), and 
angular power spectra
of difference maps of $(b34 + b45) - (b25+b33)$ (diamonds)
and of $b34 - (b45+b25)$ (circles), both with a pixelization of 3 
arcminutes. }
\label{fig:three_vs_eight}
\end{figure}

We have chosen to use both approaches.  For our high $\ell$ spectra we
limited the analysis to those 8 arcminute pixels that had more than
100 samples and for which the variation in the number of observations
in each quadrant of the pixel was less than $10$\%. Because of the
\maxima-1 scan pattern this choice corresponded to a spatial cut on
the map where the 'central section' of the map was included and the
edges excluded; the full map and the demarcation of the cut section
are shown in figure~\ref{fig:3arcm_map}. We also chose the pixel size
to be 3 arcminutes so that the effect of extra smoothing at $\ell \sim
1000$ was less than 3\%, and clearly sub-dominant compared to the
other statistical and systematic uncertainties.  The left panel in
Fig.~\ref{fig:three_vs_eight} shows the power spectra of the entire
\maxima-1 map pixelized with 8 arcminute pixels and with a
deconvolution of an approximate pixel window function (as discussed in
Section~\ref{sec:m1_sys}), only the central section pixelized with 8
arcminute pixels and with a deconvolution of the same window function,
and the entire map pixelized with 3 arcminute pixels but with no
deconvolution of a pixel window function. The conclusions are that the
spectrum at $\ell\simgt 800$ is sensitive to the details of the
pixelization, and that the 8 arcminute pixelization overestimates the
power at this $\ell$ range. Using only the well sampled 8 arcminute
pixels reduces the discrepancy between power spectra from maps with an
8 and 3 arcminute pixelizations. The 3 arcminute power spectrum shown
in the left panel of Figure~\ref{fig:three_vs_eight}, for which we used the  
data of all four detectors, can be compared with the spectra
shown in the right panel, which do include the
deconvolution of an approximate (3 arcminute) window function, and use
only the well sampled parts of the map (note that the binning at high
$\ell$ is somewhat different between the two spectra).  

\subsubsection{Data Cut}

The major parasitic signal in the \maxima-1 time stream was related to
the primary mirror modulation \cite{hanany_etal00}.  The amplitude of
this signal, which was less than $\sim 100$ $\mu$K for $b34$, $b45$
and $b25$, was comparable to the CMB signal, and therefore had to be
removed. For $b33$ the amplitude of the primary mirror synchronous
signal was $\sim 300\mu$ K and the noise inherent to this
determination was larger than for the 150 GHz detectors.  This higher
amplitude and noise were inconsequential for the determination of the
power spectrum at $\ell \simlt 800$, as has been verified extensively
in simulations and in various systematics tests (some of which have
been presented earlier in this paper).  However the effects of the
synchronous signal for $b33$ appeared non-negligible for the higher
$\ell$ regime of the power spectrum.  The power spectrum of a map made
from data that included $b33$ gave somewhat higher power at $\ell \geq
800$ compared with the power spectrum that excluded $b33$, see the
right panel of Figure~\ref{fig:three_vs_eight}.  No such difference
was found when we excluded the data from any other photometer. Power
spectra of difference maps of pairs of photometers that included $b33$
showed small inconsistencies with a null spectrum (again at $\ell \geq
800$), but power spectra of difference maps of other pairs of
photometers showed no such inconsistency. These inconsistencies were
small - for example, they essentially disappeared in the difference
maps made from combination of several photometers that included or
excluded $b33$, see the right panel of Figure~\ref{fig:three_vs_eight}
- and their origin appeared to be the mirror synchronous signal.
Foreground contributions in the \maxima-1 region were sufficiently
small and could not account for the observed inconsistencies. We
therefore chose to exclude $b33$ from the determination of the high
$\ell$ spectrum.

\section{MAXIMA-2}
\label{sec:maxima2}

The 225 square degrees area of the sky that was scanned during the
\maxima-2 flight in 1999 overlapped with 50 square degrees of the 
area scanned during
\maxima-1 and was larger by about a factor of two, 
see Figure~\ref{fig:max12scans}.  The expected detector performance and
scan strategy were similar for the two flights. 
However, the data showed a
somewhat higher level of systematic errors, which would have required
more effort to understand and overcome. The collaboration decided
to release only limited results that will facilitate the comparison
between the \maxima-1 and \maxima-2 maps.

Similar to the data from \maxima-1, it was advantageous to analyze the
\maxima-2 data that came from a subset of some of the most sensitive
photometers. Those were $b34$, $b35$, $b45$ and $b25$ operating at 
150~GHz. 
The operational parameters for these detectors including time constant, 
NET, band widths, and beam sizes are given 
in a paper by Rabii et al. \cite{inst03}.  Two of the 150 GHz
detectors gave an NET of $\sim 80\, \mu K\sqrt{s}$
and the NET for the combination of the \maxima-2 
detectors was $43 \mu K\sqrt{s}$, slightly out-performing the value of $46 \mu
K\sqrt{s}$ for the four best detectors of \maxima-1. 
\begin{figure}[t] 
\centerline{ \epsfig{file=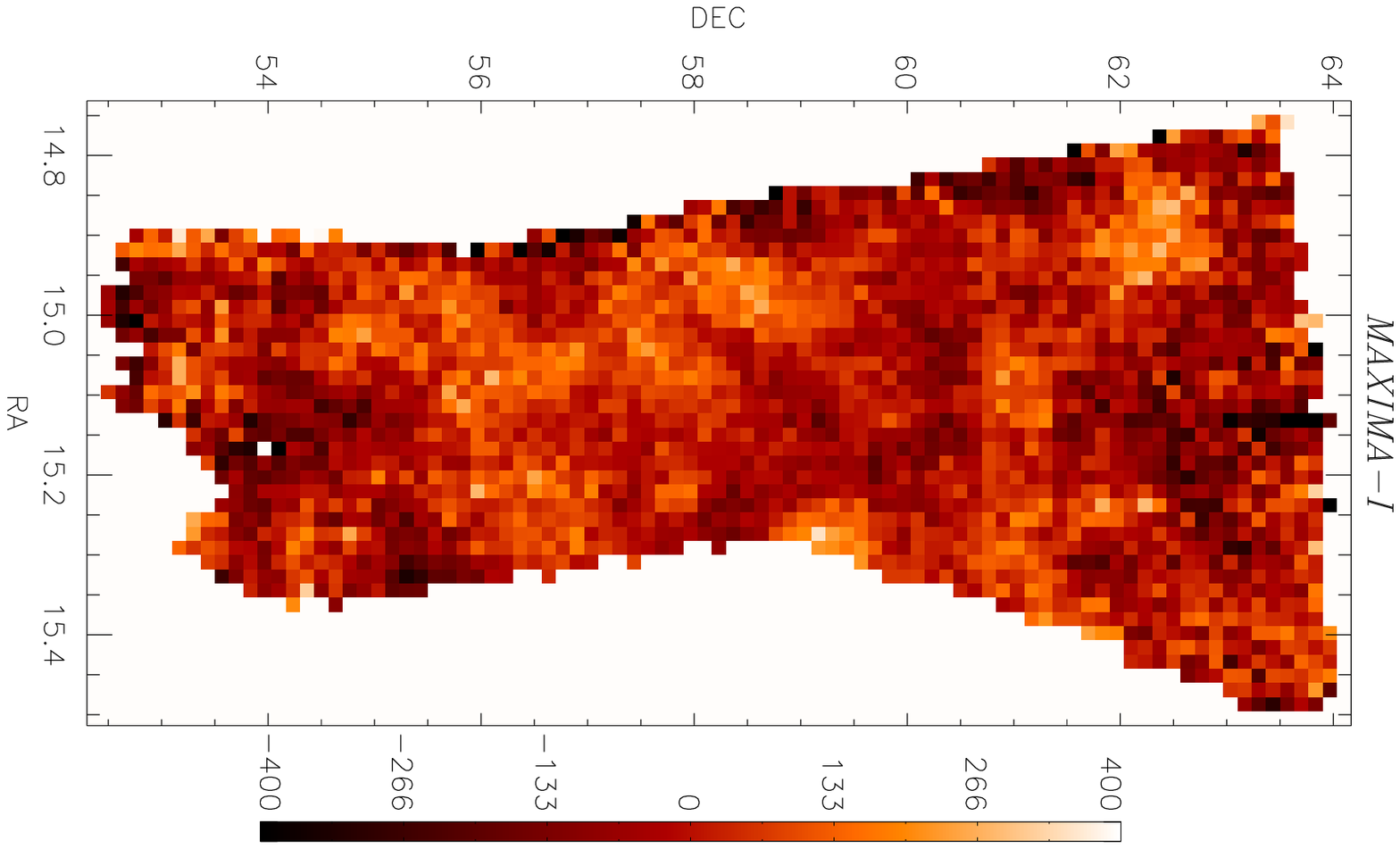,width=1.9in,angle=90}
\epsfig{file=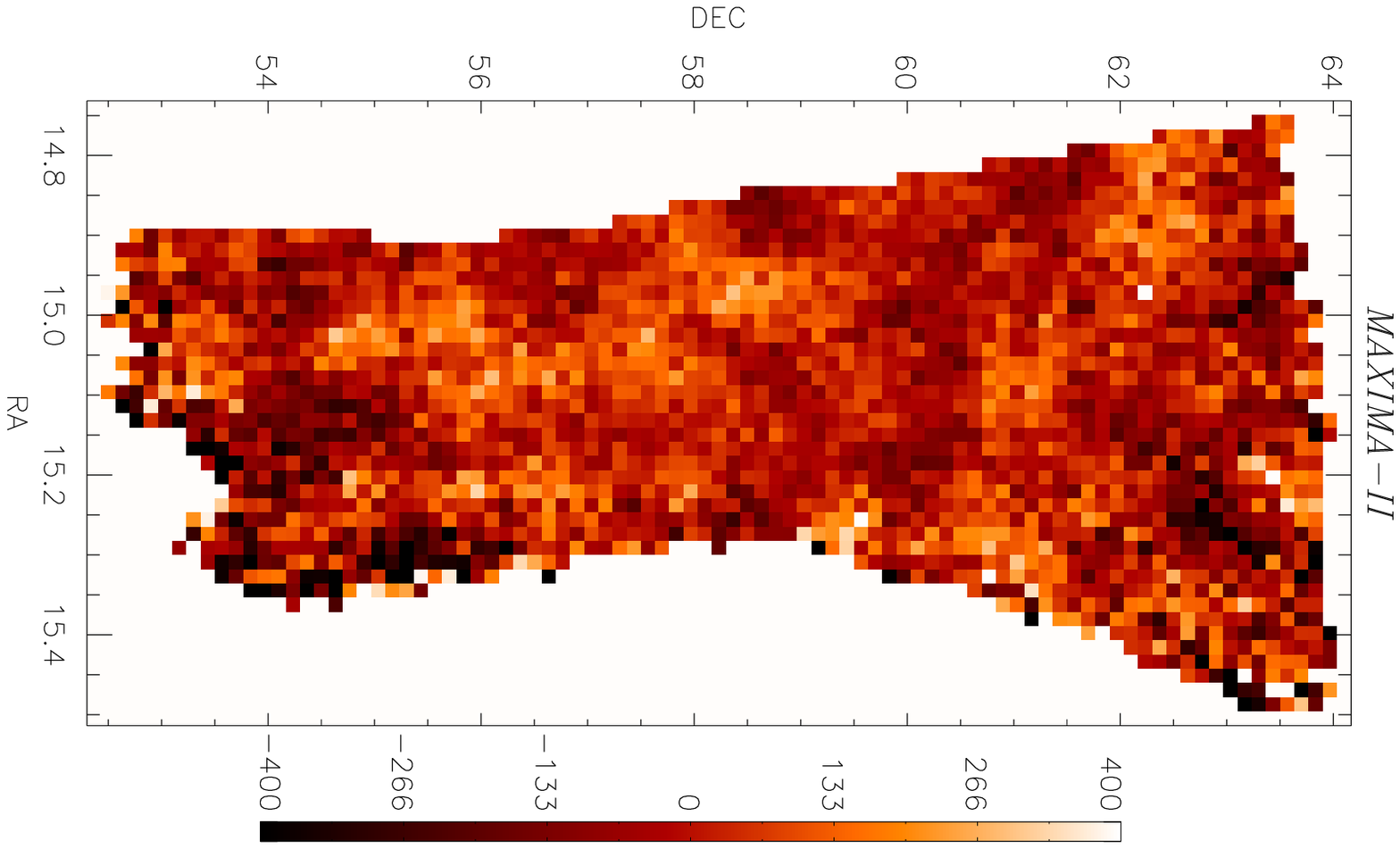,width=1.9in,angle=90}
\epsfig{file=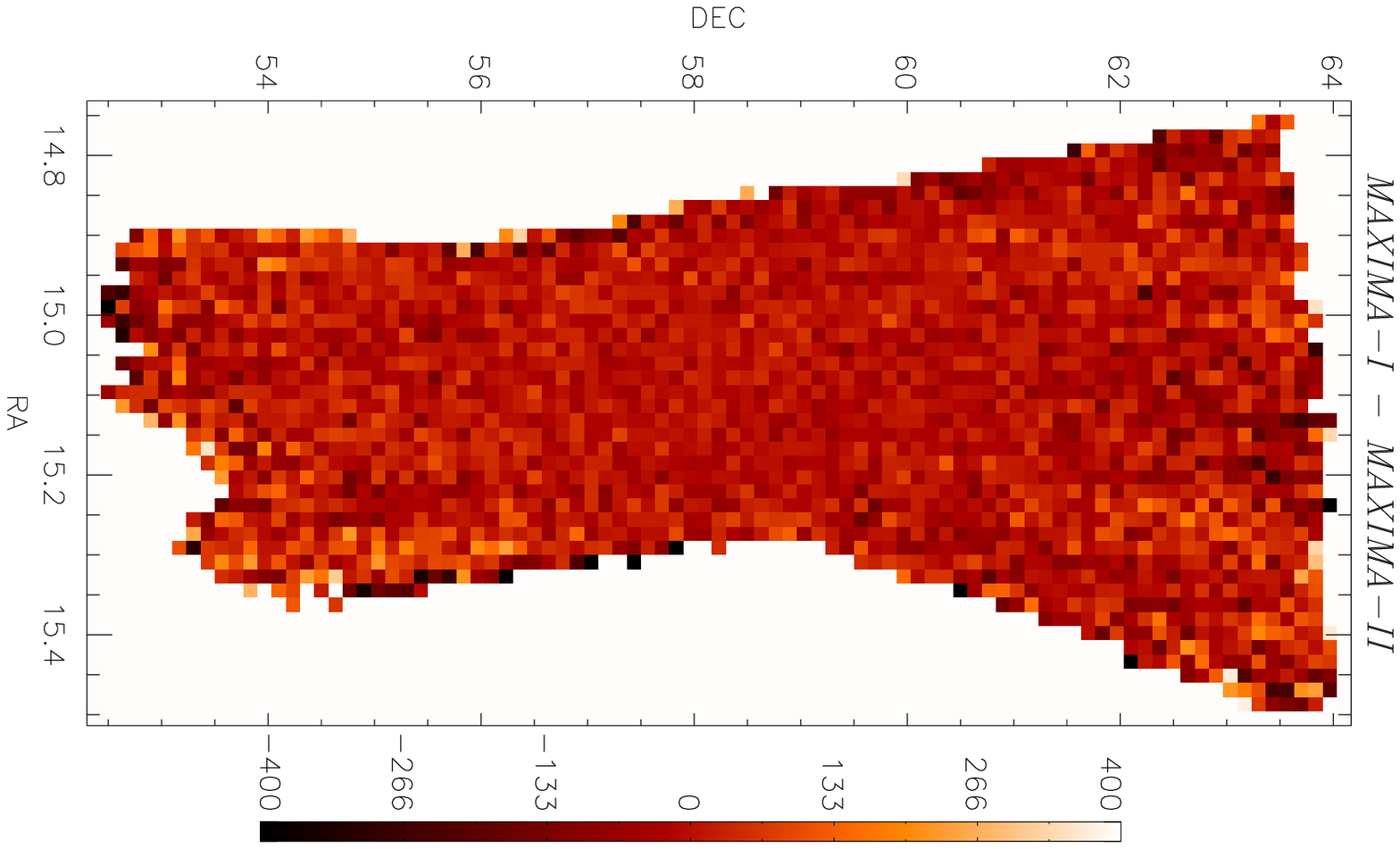,width=1.9in,angle=90} }
\caption{Maps of the overlap region between the \maxima-1 (left)
and \maxima-2 (middle) maps, and their difference (right). 
Abroe et al.~\cite{abroe_etal03} show a Wiener filtered version 
of these maps. }
\label{fig:m1m2_map}
\end{figure}

In addition to the CMB scan, the \maxima-2 flight included a
calibration on the dipole and beam mapping using Mars.  Dipole
observations in \maxima-2 were conducted at float altitude (120 kft),
unlike \maxima-1 in which they were started during ascent (70 kft).
Because of this the \maxima-2 dipole analysis did not require any
atmospheric subtraction as was done for the \maxima-1 data.  During
about 20\% of \maxima-2 CMB scan there were no detectable guide stars
for pointing reconstruction.  For this section stars were seen as
rarely as once per 30 seconds and pointing reconstruction was based on
the rate gyroscope.  The total estimated pointing error during that
time increased from 1 arcminute to 1.5 arcminutes RMS.  Other aspects
of the processing of the time ordered data, absolute calibration using
the CMB dipole, relative calibration using a mm-wave source internal
to the receiver, beam shapes determination, and pointing
reconstruction were analogous in all respects to those followed for
the \maxima-1 analysis and which are described by Hanany et al.\
\cite{hanany_etal00}.  Rabii et al. \cite{inst03} give
more details about \maxima-2. 

Estimating the maximum likelihood map also followed the prescription
given by previous publications
\cite{hanany_etal00,lee_etal01,stompor_etal02}, but the 
characteristics of the data were somewhat different than that of 
\maxima-1. There were stronger drifts giving rise to 
a $1/f^{2}$ characterization of the noise at low frequencies (compared
with $1/f$ with \maxima-1). The knee in the power spectrum 
between a $1/f^{2}$ dependence
and white noise occurred at a frequency of about
1 Hz (compared to 0.5 Hz with \maxima-1). A noise synchronous with the
modulation of the primary mirror, which has also occurred with
\maxima-1, had an amplitude of up to 500 $\mu$K, 
(as compared to less than 300 $\mu$K for
\maxima-1) and was not as stationary as in \maxima-1.
\begin{figure}[t] 
\centerline{\epsfig{file=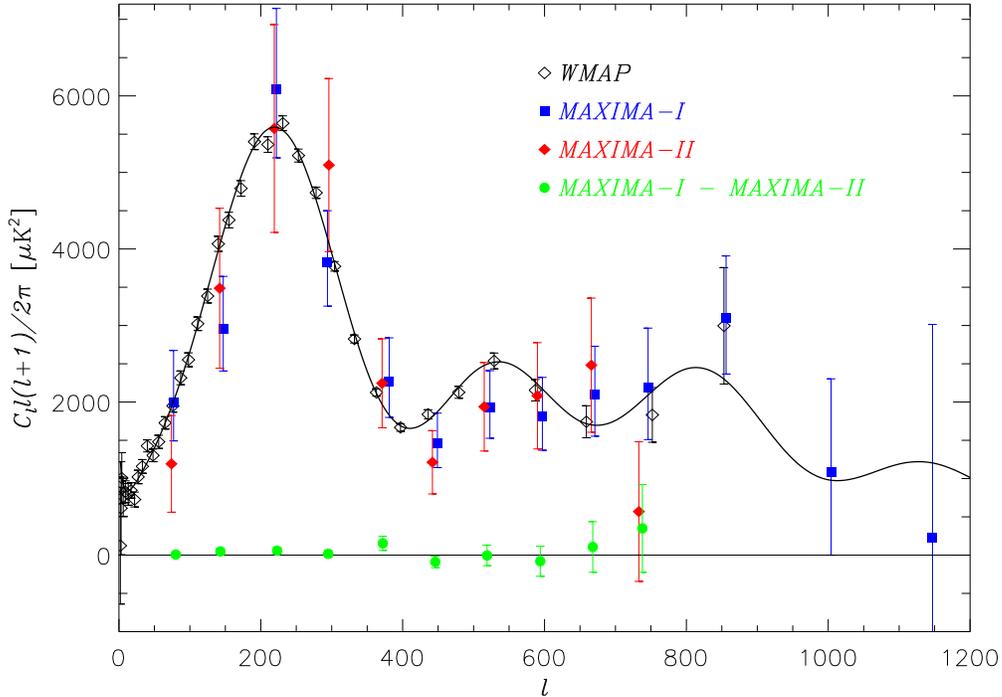,width=4.0in,angle=90}}
\caption{Power spectra from the \maxima-1 data (squares) \cite{lee_etal01}, 
the \maxima-2 data but only from 
the overlap region with \maxima-1 (filled diamonds), and a power spectrum 
of the difference map of the overlap region (circles) \cite{abroe_etal03}. 
For reference we 
also show the data from WMAP (open diamonds) and the best fit cosmology
to the WMAP data \cite{spergel_wmap}. No calibration adjustments have
been made to any of the spectra.}
\label{fig:m1m2_power}
\end{figure}

The maps made of the data of \maxima-2 and \maxima-1 in the areas
where they overlap is shown in Figure~\ref{fig:m1m2_map}. Also shown
is the difference map. To calculate the power spectra we pixelized
the maps  with 8 arcminutes square
pixels giving 5972 and 2757 pixels for \maxima-1 and -2,
respectively.  The power spectra of \maxima-2
from this overlap region, the entire \maxima-1 data \cite{hanany_etal00}, and
the spectrum of the difference map in the overlap region are shown in
Figure~\ref{fig:m1m2_power}. The $\chi^{2}$ of a null spectrum model
for the difference spectrum is 8 for 10 degrees of freedom.  Abroe et
al. \cite{abroe_etal03} have correlated this \maxima-2 map with the
maps from \maxima-1 and from WMAP 93 GHz band 
and find a high degree of correlation, providing
strong evidence that all three experiments have detected the same
spatial temperature fluctuations in this region of the sky.


\section{Summary}

The \maxima\ results, together with other CMB results
of that era, have radically changed cosmology.  The combined
COBE-DMR and \maxima\ results have constrained the flatness
of the universe and the spectral index of the power spectrum of spatial
fluctuations $n$ to unprecedented accuracy 
\cite{balbi_etal00,jaffe_etal01}
and were consistent with data from \boom\ and DASI that showed peaks
in the power spectrum at $\ell > 250$.  All of these advances together
with other astrophysical data established the current model of cosmology:
a flat universe that is overwhelmingly dominated by unknown forms of
matter and energy.
 
In this paper we presented a subset of the systematic tests that
were carried out on the \maxima-1 data before their release. We 
showed that systematic errors contributed negligibly to the
final results thereby providing the necessary confidence for the 
cosmological interpretation of the data. More recently,
the data have passed an even more stringent systematic test: 
comparison with independent data sets. The initial agreement of the 
power spectrum between \maxima-1, \boom, DASI and other experiments was 
reassuring, but the later maps of \maxima-2 (and WMAP, as shown 
by Abroe et al \cite{abroe_etal03}) give strong confidence that 
\maxima-1 has accurately mapped the cosmic microwave background 
anisotropy.


\Acknowledgements{We greatfully acknowledge support from NASA's
National Scientific Balloon Facility and NASA and NSF grants that have
supported the \maxima\ program over the years.  Computing resources
for data analysis have been provided by the Minnesota Supercomputing
Institute at the University of Minnesota and by the National Energy
Research Computing Center, which is supported by the Office of Science
of the U.S. Department of Energy under Contract No. DE-AC03-76SF00098.
MA and RS acknowledge support from NASA's grant S-92548-F. 
SH acknowledges a Land-McKnight Professorship at the University of
Minnesota. }


%
\end{document}